\documentclass[%
  reprint,superscriptaddress,%groupedaddress,
 amsmath,amssymb,pra]{revtex4-1}

\usepackage{physics}
\usepackage{xcolor}
\usepackage{xtab,afterpage,longtable}
\usepackage{lipsum}
\usepackage{graphicx}% Include figure files
\usepackage{dcolumn}% Align table columns on decimal point
\usepackage{booktabs} % For professional looking tables
\usepackage{multirow}
\usepackage{color}
\newcommand*{\rom}[1]{\uppercase\expandafter{\romannumeral #1}}

\usepackage{hyperref}% add hypertext capabilities
\usepackage{hyperref}% add hypertext capabilities
\usepackage[utf8]{inputenc}
\usepackage{makecell}
\usepackage[normalem]{ulem}
\usepackage{array}         % for m{..}, >{..} etc.
\usepackage{booktabs}      % optional; nicer rules
\newcommand{\DeltaT}{\Delta\Gamma_1}

\begin{document}

\title{Quantum Sensing of Copper-Phthalocyanine Electron Spins via NV Relaxometry}% Force line breaks with \\

\author{Boning Li}\thanks{These authors contributed equally.}
    \affiliation{Department of Physics, Massachusetts Institute of Technology, Cambridge, MA 02139, USA}
    \affiliation{Research Laboratory of Electronics, Massachusetts Institute of Technology, Cambridge, MA 02139, USA}
\author{Xufan Li}\thanks{These authors contributed equally.}
   \affiliation{Honda Research Institute USA, Inc., San Jose, CA 95134, USA}
\author{Yifan Quan}
    \affiliation{Department of Chemistry and Francis Bitter Magnet Laboratory, Massachusetts Institute of Technology, Cambridge, MA 02139, USA}
\author{Avetik R Harutyunyan}\email{aharutyunyan@honda-ri.com}
    \affiliation{Honda Research Institute USA, Inc., San Jose, CA 95134, USA}
    \affiliation{Department of Nuclear Science and Engineering, Massachusetts Institute of Technology, Cambridge, MA 02139, USA}
\author{Paola Cappellaro}\email[]{pcappell@mit.edu}
    \affiliation{Department of Physics, Massachusetts Institute of Technology, Cambridge, MA 02139, USA}
    \affiliation{Research Laboratory of Electronics, Massachusetts Institute of Technology, Cambridge, MA 02139, USA}
    \affiliation{Department of Nuclear Science and Engineering, Massachusetts Institute of Technology, Cambridge, MA 02139, USA}

\begin{abstract}
Molecular spin systems are promising candidates for quantum information processing and nanoscale sensing, yet their characterization at room temperature remains challenging due to fast spin decoherence. In this work, we use $T_1$ relaxometry of shallow nitrogen-vacancy (NV) centers in diamond to probe the electron spin ensemble of a polycrystalline copper phthalocyanine (CuPc) thin film. In addition to unequivocally identifying the  NV-CuPc interaction thanks to its hyperfine spectrum, we further extract key parameters of the CuPc spin ensemble, including its correlation time and local lattice orientation, that cannot be measured in bulk electron resonance experiments. The analysis of our experimental results  confirms that electron-electron interactions dominate the decoherence dynamics of CuPc at room temperature. Additionally, we demonstrate that the CuPc-enhanced NV relaxometry can serve as a robust method to estimate the NV depth with $\sim1$~nm precision.
Our results establish NV centers as powerful probes for molecular spin systems, providing insights into molecular qubits, spin bath engineering, and hybrid quantum materials, and offering a potential pathway toward their applications such as molecular-scale quantum processors and spin-based quantum networks.

\end{abstract}
%PC: Just a radom idea, not to include in the paper: we could have also maybe measured the CuPc layer thickness? (by using protons before/after burn). Maybe 20nm is too much, but if we do a thinner layer... it could be a method to characterize thin film deposition!
\maketitle
\section{Introduction}
Quantum technology promises to provide unprecedented capabilities in computation, sensing and communication~\cite{steane1998quantum,degen2017quantum,cai2013large,tang2023communication,li2024blind}. Reaching the expected performances requires building robust quantum platforms that respond to the needs of the information task at hand. Among different quantum systems that have been proposed, electronic and nuclear spins have emerged as the platform of choice for task-oriented computations (such as simulations) ~\cite{abobeih2022fault,godfrin2017operating,ruskuc2022nuclear,lim2025demonstrating} and for quantum sensing~\cite{qiu2021nuclear,wang2024hyperfine}. Over the past decade, paramagnetic molecules with electron and nuclear spin have gained increasing attention as promising quantum platforms because of their robustness under complex condition and great tunability through chemical modifications and crystal engineering~\cite{atzori2016room,mullin2024systems,warner2013potential,wedge2012chemical,lavroff2021recent,gaita2019molecular}. Exploiting the advantages of the rich spin states in molecules~\cite{godfrin2017operating} requires overcoming
their limitations, particularly the difficulty in initialization and readout. 

Conventional electron paramagnetic resonance (EPR) techniques rely on thermal equilibrium for spin initialization, typically occurring on millisecond timescales~\cite{warner2013potential}; the spin state is read out via magnetization which requires an ensemble of spin qubits. Recent progress in engineering the spin-photon interaction of certain molecule demonstrated spin-select photon absorption and emission processes that offer the potential for efficient optical pumping to initialize and read-out~\cite{bayliss2020optically}. Not only this is only achievable on certain molecules but it requires  milliKelvin temperatures. Here we propose and take the first steps to demonstrate an alternative method to achieve  local and fast control of the spins in a molecule in a broader range of conditions, by exploiting their interactions with Nitrogen vacancy (NV) centers in diamond. The NV center is known for its excellent spin-optical properties enabling fast single-spin initialization and read-out and long coherence at ambient conditions ~\cite{doherty2013nitrogen}. Additionally, the NV center electron spin state   is highly sensitive the external electro-magnetic field and it has been used to perform noise spectroscopy of its environment~\cite{pham2016nmr,sangtawesin2019origins,sun2022self,wang2024digital}.
 NV centers have also demonstrated coherent coupling to isolated electronic spins~\cite{sushkov2014magnetic,cooper2020identification,knowles2016demonstration,rosenfeld2018sensing,degen2021entanglement}. The  polarization and coherence transfer protocols recently deployed to manipulate environmental spin qubits via dipolar interaction~\cite{ungar2024control}  are transferable to the hybrid NV-molecule spin systems.

To demonstrate the feasibility of such hybrid system, here we probe the interaction between a single NV center in diamond and the electron spins of copper phthalocyanine (CuPc)~molecules at room temperature. CuPc is a well-known  material with well-developed techniques to synthesize, fabricate and deposit~\cite{warner2013potential,topuz2013synthesis,cranston2021metal}. 
Electron paramagnetic resonance studies of the electron spin from the copper atom d-orbital have highlighted its long coherence time at cryogenic temperatures. The electron spin is strongly coupled with copper ($^{63}$Cu and $^{65}$Cu) and nearby nitrogen $^{14}$N nuclear spins,  that could also be used as quantum registers. Quantum protocols using the electron spin to polarize and manipulate the nuclear spin state for high performance quantum computing are established in similar systems~\cite{ruskuc2022nuclear,godfrin2017operating}.

 We quantitatively demonstrate the interaction between NV center with CuPc electron spin with  T$_1$ relaxometry experiments~\cite{steinert2013magnetic, pelliccione2014two, kumar2024room, li2019all} and develop an accurate model  that enable extracting key properties  of the CuPc spin system from the experimental results. We first demonstrate the role of the hyperfine spectrum, a first step towards exploiting the role of the long-lived nuclear spins for quantum applications.
We further introduce the explicit form of the interaction strength resulting from our theoretical model and how its fit to our experimental data  reveals properties of the CuPc, including the spin bath  correlation time and nano-scale variations of the thin film lattice orientation that cannot be measured on bulk materials. Our results further identify electron-electron spin interactions as the dominant mechanism for the CuPc electronic spin decoherence. 
%  The effect of the CuPc electron spin bath as an effective fluctuating magnetic field acting on the NV spin. The resulting NV spin depolarization is then predicted using Fermi's golden rule. 
% We derive the explicit form of the interaction strength characterization, and successfully used it to fit the experimental depolarization data. Key parameters of the spin bath including correlation time and lattice orientation of CuPc thin film in nanometer scale. We subsequently studied the decoherence mechanism and proves it is from electron-electron spin interaction. 
Finally, thanks to our quantitative understanding of the interaction strength, we propos a novel method to measure the depth of shallow, single NV centers, demonstrating its advantages over existing techniques~\cite{pham2016nmr,grinolds2014subnanometre}.

\section{Experimental Conditions and Theoretical Model}
\subsection{NV center and CuPc thin film condition}
To engineer our hybrid NV-CuPc system, we deposited 
a thin film of pure CuPc (27 $\pm$0.8~nm thickness)    on the diamond surface~\cite{cranston2021metal}. The crystal phase of the CuPc thin film was identified as the $\alpha$-phase, based on characteristic Raman peaks~\cite{shaibat2010distinguishing}.  
We further measured photoluminescence spectra of the NV-CuPc sample under 532~nm excitation. The emission spectra from the NV centers and CuPc were spectrally distinguishable, enabling selective detection of NV fluorescence using appropriate filtering.
Using a home-built confocal setup, we  evaluated the stability of the CuPc thin film under the 532~nm laser illumination required for NV detection, and determined the optimal laser power and exposure duration that preserved the integrity of the CuPc layer. Applying appropriate filtering to collect light in the 594--715~nm range, we then demonstrated fluorescence detection of individual NV centers beneath the CuPc layer.

To demonstrate NV-to-CuPc interaction, we used a diamond containing shallow NV centers created via low-energy ion implantation followed by annealing, with an estimated depth of approximately 10-20~nm
%14 nm 
from the surface.  Prior to thin-film deposition, the diamond surface was confirmed to be free of electron spins through double electron-electron resonance (DEER) measurements.  Throughout the experiments, a static magnetic field was applied with precise alignment along the NV axis. 

While normal laser excitation allowed typical NV experiments (ODMR, spin echo, etc.),  higher laser power and prolonged illumination caused localized degradation of the CuPc film. This effect enables direct comparative measurements on the same NV center, both with and without the CuPc layer, allowing us to isolate the influence of CuPc on the NV spin properties. 

Additional details on material properties and laser power characterization are provided in the Supplementary Material.

\begin{figure}[htbp]
\centering
\includegraphics[width=\linewidth]{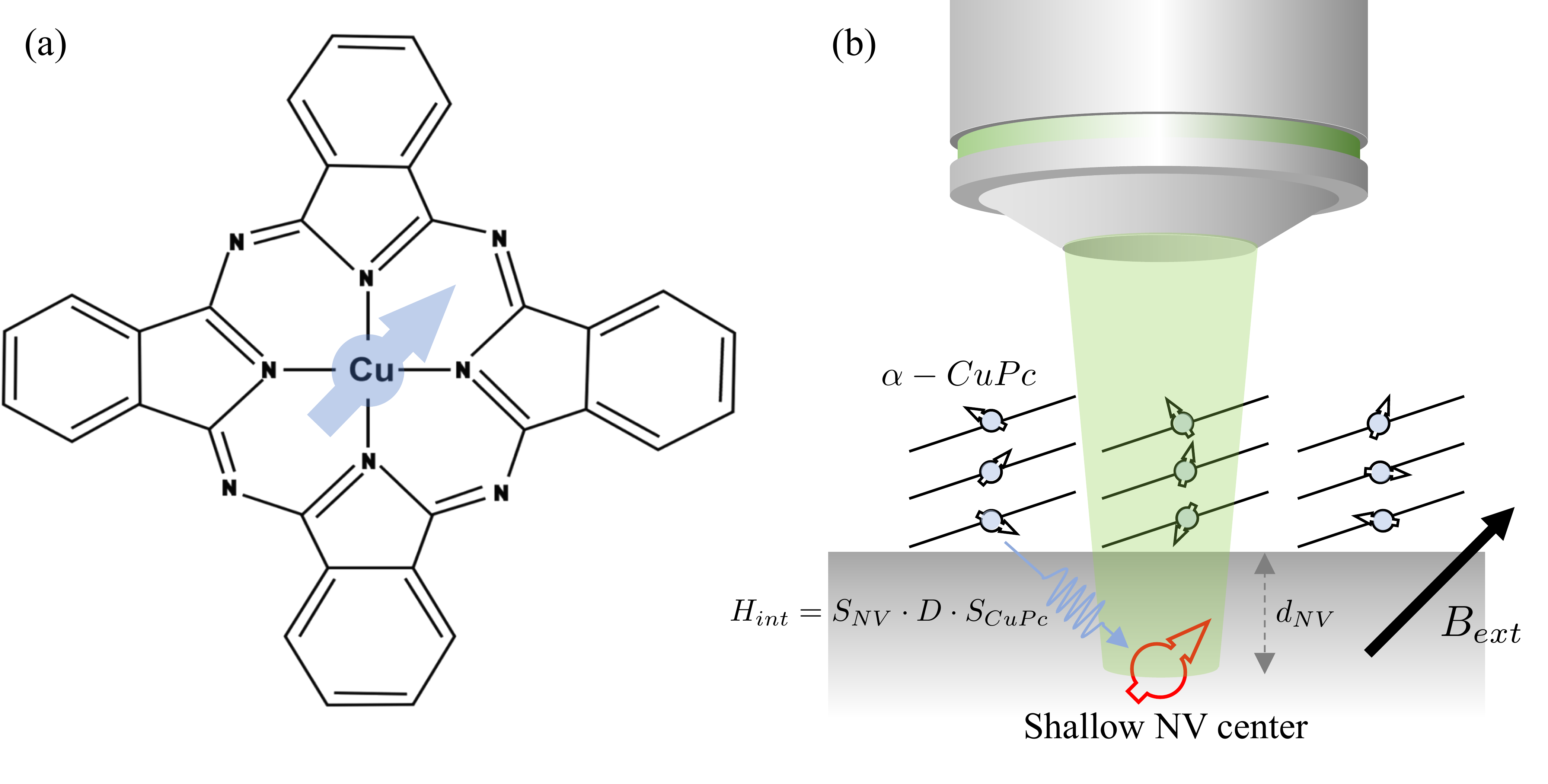}
\caption{Experimental Setup.
(a) Schematic of the experiment, where a shallow NV center in diamond is initialized and read out using a 532~nm laser. The NV center couples to the electron spin of CuPc molecules via magnetic dipolar interaction.  
(b) Illustration of the CuPc thin film deposited on the diamond surface. The red arrow represents the NV center in a (100)-oriented diamond, while the blue shaded arrows indicate the non-polarized electron spins of CuPc. An external magnetic field is applied along the NV axis.  
}
\label{fig:experiment_condition}
\end{figure}

\subsection{Modeling the interaction between the NV center and CuPc electron spin ensemble}
% I would flip it around: 1. spectrum (as done in the past, add references  https://journals-aps-org.libproxy.mit.edu/prapplied/abstract/10.1103/PhysRevApplied.2.054014  https://pubs-acs-org.libproxy.mit.edu/doi/10.1021/acs.nanolett.9b02960 https://pubs.acs.org/doi/10.1021/acs.nanolett.3c05090) 
%https://www-nature-com.libproxy.mit.edu/articles/ncomms2588
%2. spectrum calculated from autocorrelation, quite typically taken as OU model or Lorentzian spectrum. However, due to the complexity of the CuPc spectrum, we need to more carefully analyze the autocorrelation. Our modeling provides a dependence on the hyperfine strength, on the CuPc correlation time, and on the angular orientation of the molecule that we can compare to experimental results. 
%This presentation has too many details (but it's then still redone in the appendix).  I think we can say (1) (sorry about the order) that the noise strength can be calculated from a uniform field (b0 equation). (2) that we expect a contribution at zero frequency and sum of contributions at all hyperfine transition energies (still with a Lorentzian spectrum).
%So keep only eq. 8 and 9 + defin what eta are (and relate to appendix).
The effects of a spin environment such as the CuPc on an isolated spin probe can be modeled as
 an effective random magnetic field.  In particular, this magnetic field transverse  component drives transitions between the NV center $\ket{0}$ and $\ket{\pm1}$ states, which results in a longitudinal relaxation with relaxation time $T_1$~\cite{choi2017depolarization, machado2023quantum,steinert2013magnetic, pelliccione2014two, kumar2024room, li2019all}. NV $T_1$ relaxometry has been employed to characterizes magnetic noise from electron spin baths in various systems~\cite{steinert2013magnetic, pelliccione2014two, kumar2024room, li2019all}. Indeed, $T_1$ probes the spin bath noise spectrum $S_e$ at the NV frequency,
~\cite{choi2017depolarization, slichter2013principles}:
\begin{equation}
\frac{1}{T_1} = {\gamma_e^2}  S_e(\omega_{\text{NV}}),
\label{eq:T1_mech}
\end{equation}
where $\omega_{\text{NV}}$ is the NV transition frequency between $\ket{0}$ and $\ket{-1}$.
In previous studies, $S_e(\omega)$ was typically taken as 
%bath is often modeled as an Ornstein–Uhlenbeck (OU) magnetic field noise source, with 
a Lorentzian spectral density centered at the free electron Larmor frequency.  The strong hyperfine coupling of CuPc to nuclear spins requires a more detailed treatment of the spin environment dynamics and its coupling to the NV center. Interestingly, our analysis reveals that the resulting $T_1$ depends sensitively on (and can thus characterize)  the hyperfine structure, the spin correlation time $\tau_e$, and the depth of the NV center relative to the CuPc layer.

The approximation of the CuPc as a classical  noise source is justified since the  interaction with a single  NV center does not produce any quantum backaction since it is much weaker than internal interactions. Indeed,    the dipolar coupling of an NV located approximately 10~nm below the diamond surface with a CuPc spin   $|D| \sim 2\pi\times50~\text{kHz}$, is negligible compared to the intrinsic decoherence rate of CuPc electron spins ($>2\pi\times4.2~\text{MHz}$~\cite{smofthispaper}).

The stochastic transverse magnetic field $B_{\perp}$ at the NV site arises from dipolar interactions with all CuPc spins in the thin film,
\begin{equation}
    B_{\mu} = \sum_{n} \sum_{\nu = x,y} D_{\mu\nu}^{n} S_{\nu}^{n}, \quad \mu = x, y,
    \label{eq:B_main}
\end{equation}
where $n$ indexes CuPc molecules and $D_{\mu\nu}^{n}$ is the dipolar coupling strength between the $n^{th}$ CuPc spin and the NV center.

The noise spectrum $S_e$ is the Fourier transform of  the field autocorrelation function~\cite{choi2017depolarization,machado2023quantum,rodriguez2018probing}:
\begin{equation}
    G_e(t) = \langle B_{\perp}(0) B_{\perp}(t) \rangle.
    \label{eq:G0}
\end{equation}

% This field is constructed from the sum over dipolar contributions from individual CuPc spins:
%  Since we are concerned with longitudinal relaxation, we only consider the transverse field components ($\mu = x,y$). The stochastic fluctuations due to fast-decaying, non-polarized CuPc spins yield:
% \begin{equation}
%     G_e(t) \sim \sum_{n,m} D_{\mu\nu}^{n} D_{\mu\nu'}^{m} \langle S_{\nu}^{n}(0) S_{\nu'}^{m}(t) \rangle,
% \end{equation}
% where $\langle S_{\nu}^{n}(0) S_{\nu'}^{m}(t) \rangle$ is the CuPc spin correlation function.

A detailed analysis of CuPc spin dynamics (see  Appendix.~\ref{app:auto_correlation_function})  shows that their transverse spin autocorrelation contains both a quasi-static contribution from CuPc's electron spin longitudinal relaxation and a set of discrete frequency components associated with hyperfine transitions. By introducing the correlation time $\tau_e$ for the electronic spin states of CuPc, we obtain:
\begin{equation}
    G_e(t) = b_0^2 e^{-t/\tau_e} \left[ \frac{5}{16} + \frac{11}{16} \sum_{i,j} \eta_{ij} \cos(\omega_{ij} t) \right],
\end{equation}
where $b_0^2=\langle B_\perp(0)^2\rangle$. Here $i$ and $j$ label different hyperfine eigenstates $\ket{\psi_i}$ and $\ket{\psi_j}$, with energy separation $\hbar\omega_{ij}$. A weight factor $\eta_{ij} = \frac{|\langle \psi_i | S_{\perp} | \psi_j \rangle|^2}{M}$ is added to each transition pairs ($i,j$), where $S_{\perp}$ is the transverse electron spin operator  and $2M$ is the total number of hyperfine states. This factor captures both the transition amplitude between the hyperfine states and the near-equal thermal population of states at room temperature.

Assuming all CuPc molecules are identical and their distance much smaller than the NV depth, the total coupling strength can be evaluated by integrating over a uniform thin film of thickness $h$ and spin density $n_e$;
\begin{equation}
    b_0^2 = 
\left( \frac{\mu_0 \hbar \gamma_{e}}{4\pi} \right)^2 \frac{2\pi S_e(S_e+1)}{9} n_e \left( \frac{1}{d_{\text{NV}}^3} - \frac{1}{(d_{\text{NV}} + h)^3} \right),
\end{equation}
where $\mu_0$ is the vacuum permeability, $\gamma_e$ the electron gyromagnetic ratio, $S_e(S_e+1) = 3/4$ for a spin-1/2 electron, and $d_{\text{NV}}$ is the NV center’s depth.

The Fourier transform of $G_e(t)$ yields the power spectral density:
\begin{align}
    S_e(\omega) &= 
    \frac{5}{8}  \frac{b_0^2 \tau_e}{\omega^2 \tau_e^2 + 1} \label{eq:spec} \\
    &+ \frac{11}{16} \sum_{i,j} \eta_{ij} \left[ \frac{b_0^2 \tau_e}{(\omega_{ij} - \omega)^2 \tau_e^2 + 1} 
    + \frac{b_0^2 \tau_e}{(\omega_{ij} + \omega)^2 \tau_e^2 + 1} \right].\nonumber
    \label{eq:spectra_density}
\end{align}
%A detailed derivation with consideration of isotopes of nuclei was shown in Appendix.~\ref{app:auto_correlation_function}. NV center’s longitudinal relaxation rate due to this fluctuating magnetic field is given by Fermi’s golden rule
This model predicts that the NV $T_1$ time (Eq.~\ref{eq:T1_mech}) depends strongly on the NV depth $d_{\text{NV}}$, the CuPc spin correlation time $\tau_e$, and the hyperfine transition spectrum $\omega_{ij}$, which in turn varies with the molecular orientation $\theta_e$ relative to the NV quantization axis due to the strong anisotropic hyperfine interaction in CuPc~\cite{finazzo2006matrix}. In the following sections, we apply this model to extract these parameters from the experimental data.

\begin{figure}[htbp]
\centering
\includegraphics[width=\linewidth]{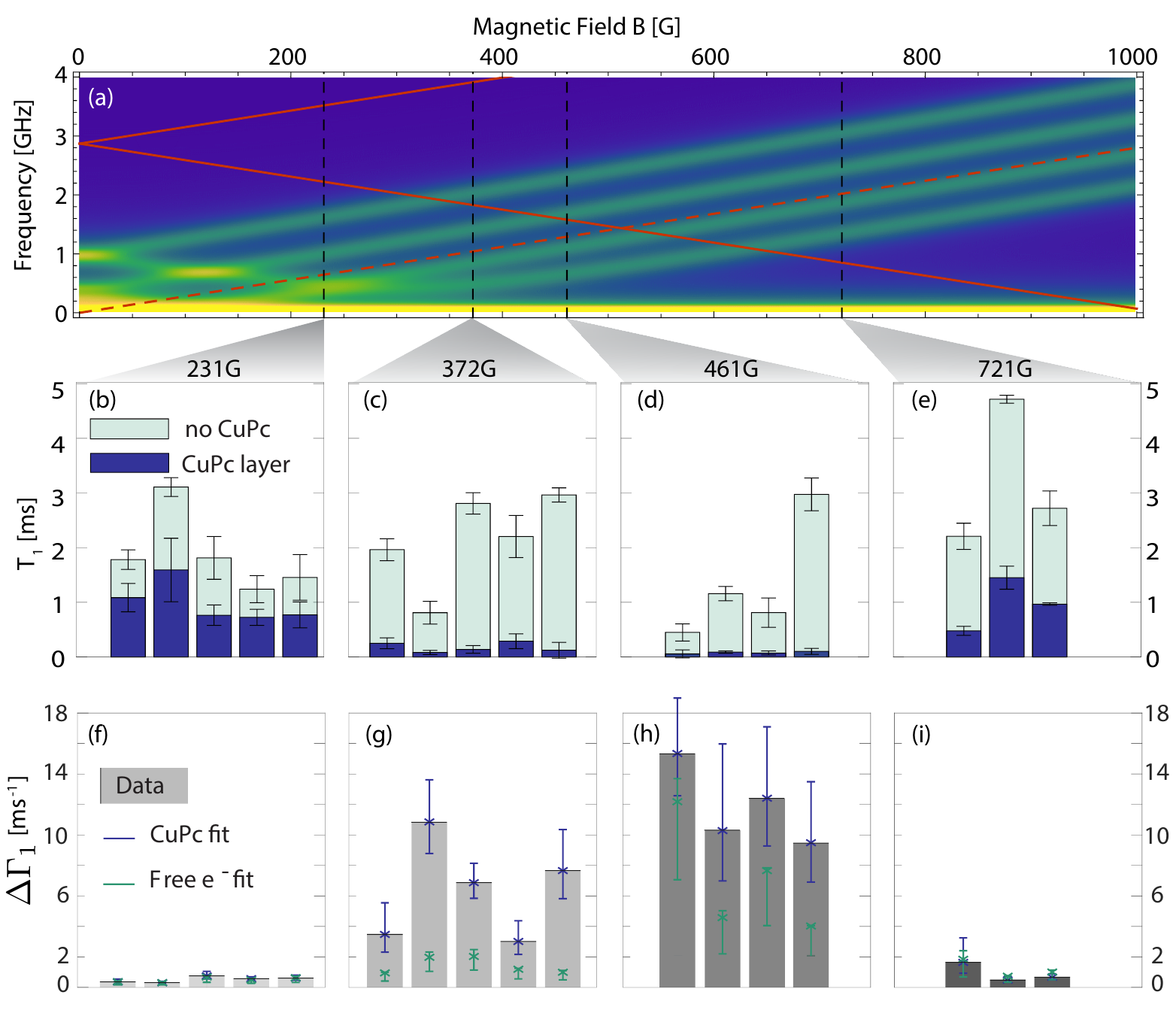}
\caption{\textbf{NV Relaxometry}.  
(a) Simulated CuPc spectral density $S_e$ (broad line),  NV  resonance frequency ($\ket{0} \rightarrow \ket{\pm1}$ transition, red solid line) and free electron spin resonance frequency   (red dashed line) as a function of external magnetic field strength. The four split peaks correspond to the hyperfine states of the CuPc electron spin, arising from coupling to the copper nuclear spin. They are broaden due to hyperfine with nitrogen nuclear spins.  Vertical gray dashed lines indicate the four magnetic fields selected for experimental measurements.  
(b-e) $T_1$  of individual NV centers at the four magnetic fields, both in the presence (blue) and absence  of the CuPc film.  
(f-i) Change in NV center depolarization rate, $\DeltaT$ due to the presence of CuPc. The two fitting results correspond to different assumptions about the source of the electron spin bath, with both assuming that the electron spatial density equals the CuPc molecular density. 
}
\label{fig:T1_measurement}
\end{figure}

\section{CuPc Spectroscopy and Characterization}
To probe the interaction between the NV center and the electron spins of CuPc, we performed $T_1$ relaxometry measurements on multiple single NV centers under varying magnetic fields. The measurements were first conducted in the presence of the CuPc layer, yielding the depolarization time  $[T_1]_{\text{CuPc}}$. Then, the CuPc layer was removed using high-power laser illumination, and a second relaxation measurement was performed on the same NV centers to obtain the depolarization time in the absence of external electron spins, denoted as $[T_1]_{\text{free}}$.

As shown in Fig.~\ref{fig:T1_measurement}, not only we see a change in $T_1$ times when removing the CuPc layer, but the difference depends on the external magnetic field. To understand this behavior, we plot in Fig.~\ref{fig:T1_measurement}(a)  the spectral density of the CuPc spin bath, $S_e(\omega)$ (Eq.~\eqref{eq:spec}) and compare it with the NV center’s $\ket{0} \rightarrow \ket{\pm1}$ transition frequency as a function of magnetic field. When the magnetic field brings the NV transition energy into resonance with the CuPc spectrum, enhanced polarization transfer occurs between the two spin systems, leading to a reduction in $T_1$, as predicted by Eq.~\eqref{eq:T1_mech},.

To quantify the CuPc-induced depolarization, we define the change in the NV relaxation rate,
\begin{equation}
\DeltaT^{\text{exp}} = \left[\frac{1}{T_1}\right]_{\text{CuPc}} - \left[\frac{1}{T_1}\right]_{\text{free}},
\end{equation}
and fit the data to our theoretical model in Fig.\ref{fig:T1_measurement}(c). The results  clearly show a significant reduction in $T_1$ in the presence of the CuPc layer, particularly at magnetic fields near 372 G and 461 G where the NV and CuPc frequencies overlap. 

With the  CuPc layer thickness $h$, spin density $n_e$,   and NV depth $d_{\text{NV}}$\cite{pham2016nmr}  independently determined, we fit the experimental data using our theoretical model [Eqs.~(\ref{eq:T1_mech},~\ref{eq:spec})] to extract information about the spin bath. The resulting fit shows good agreement with the experimental measurement ( the black solid lines in Fig.\ref{fig:T1_measurement}(c)). In the following sections, we describe how  the fitting can extract  the spin bath parameters.

\subsection{Molecular orientation}
\begin{figure}[b!]
\centering
\includegraphics[width=\linewidth]{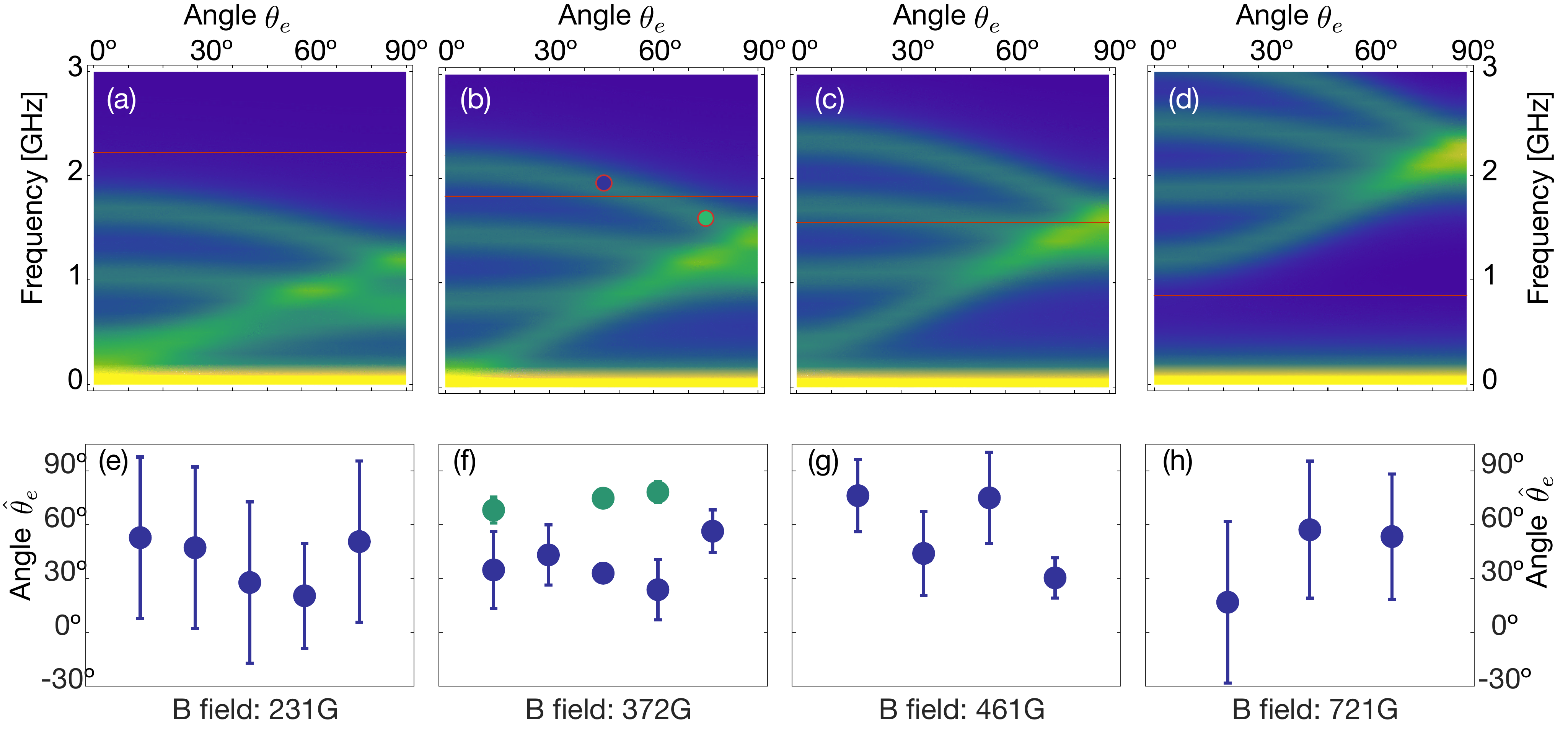}
\caption{CuPc Crystal Orientation Estimation.
(a–d) Simulated CuPc spectral density with the NV center's resonance frequency overlaid as a red solid line, as a function of the relative orientation between the CuPc molecule and the external magnetic field. (a)–(d) correspond to the four magnetic fields studied. (e–h) Estimated CuPc molecular orientations on top of individual NV centers at the corresponding magnetic fields. 
At 231~G (a, e), there is no spectral overlap between the CuPc and NV transitions across all $\theta_e$, resulting in a large uncertainty in the estimated orientation. A similar situation occurs at 721~G (d, h). We leverage this minor effect of $\theta_e$ on $S_e(\omega_{NV})$ for accurate estimation of $\tau_e$ and $d_{NV}$. In contrast, at 461~G (c, g), the degree of spectral overlap depends more sensitively on $\theta_e$, allowing for a relatively precise estimation of the orientation angle. At 372~G (b, f), a crossing between the NV center resonance and one of the CuPc hyperfine transitions occurs, leading to two possible estimated orientations, shown in blue and red.
}
\label{fig:orientation}
\end{figure}
As NV relaxometry probes the spectrum of the spin bath, it can relay information about the orientation  between the external magnetic field and the CuPc molecular plane. The orientation angle $\theta_e$ influences the spin bath transition frequencies $\omega_{ij}$, because of the anisotropy of both the $g$-factor and the hyperfine couplings in CuPc~\cite{finazzo2006matrix}. 
  A priori, we can expect that each NV center probes a distinct local environment characterized by specific values of  $\theta_e$~\cite{cranston2021metal} and correlation time $\tau_e$.  To extract these two parameters,  we  fit the experimental data to the theoretical model via nonlinear least-squares parameter estimation. Aided by the fact that the NV depth can be measured independently after removing the CuPc layer~\cite{pham2016nmr} and assuming the hyperfine interaction values reported in the literature~\cite{}, only $\lambda = \{\tau_e, \theta_e\}$ need to be determined  by minimizing the relative deviation between the theoretically predicted and experimentally measured NV depolarization rates:
% \begin{equation}
%     \hat{\lambda} = \mathop{\arg\min}\limits_{\lambda} \left\| \frac{\DeltaT^{\text{exp}} - \DeltaT^{\text{th}}(\lambda) }{ \DeltaT^{\text{exp}} }\right\|^2.
% \end{equation}
\begin{equation}
    \hat{\lambda} = \mathop{\arg\min}\limits_{\lambda} \left\|\DeltaT^{\text{exp}} \ - \DeltaT^{\text{th}}(\lambda) \right\|^2.
    \label{eq:fitting_value}
\end{equation}
We further simplify the fitting process by first considering magnetic fields 
such as 230~G and 720~G (Fig.~\ref{fig:orientation}(a, c)) 
where the NV transition frequency is sufficiently detuned from the CuPc spectral features over the whole  range of $\theta_e$. Under these conditions, $T_1$ relaxation is predominantly sensitive to the correlation time $\tau_e$ (the following result of large uncertainty in $\theta_e$ estimation at this magnetic field condition justifies this assumption). At these conditions, the two-parameter fitting of experiment data yields $\hat{\tau}_e = (2.0 \pm 1.1)$ns over all NVs. (Fitting result of $\tau_e$ for each NV center is shown in Supplementary materials).

With the correlation time $\tau_e$ determined,  we estimate the orientation angle $\theta_e$ using the same fitting procedure. To further evaluate the goodness of the fit, we estimate the parameter uncertainty taking into account both measurement statistics and the uncertainty in independently-measured parameters. We collect these parameters 
%--the CuPc layer thickness $h$, electron spin density $n_e$, and NV center depth $d_{\text{NV}}$-- 
in $\lambda_{\text{ind}} = \{d_{\text{NV}}, h, n_e\}$, and denote their  95\% confidence intervals as $I_{\text{ind}}$. The uncertainty of our estimates of $\lambda=\{\theta_e,\tau_e\}$ is then obtained as the range of $\lambda$ for which the difference from the theoretical value is less than the experimental uncertainty, $\epsilon_{\DeltaT^{\text{exp}}}$   in measuring $\DeltaT^{\text{exp}}$,
\begin{equation}
\left| \DeltaT^{\text{exp}} - \DeltaT^{\text{th}}({\lambda},{\lambda_{ind}})   \right| < \epsilon_{\DeltaT^{\text{exp}}},
\label{eq:fitting_error}
\end{equation}
evaluated over all confidence interval (${\lambda}_{ind}\in\ I_{ind}$) of the independent parameters.
% \begin{equation}
% \begin{aligned}
% \hat{\lambda}\in \left\{{\lambda} \ \middle|\ 
% \begin{array}{c}
% \left| \frac{\DeltaT^{\text{exp}} - \DeltaT^{\text{th}}({\lambda},{\lambda_{ind}}) }{ \DeltaT^{\text{exp}} }\right| < \epsilon_{([\Delta \frac{1}{T_1}]_{\text{exp}})} ,\\\quad \text{for} \ {\lambda}_{ind}\in\ I_{ind}
% \end{array}
% \right\}.
% \end{aligned}
% \end{equation}
% \begin{equation}
% \begin{aligned}
% \hat{\lambda}\in \left\{{\lambda} \ \middle|\ 
% \begin{array}{c}
% \left| \DeltaT^{\text{exp}} - \DeltaT^{\text{th}}({\lambda},{\lambda_{ind}})   \right| < \epsilon_{\DeltaT^{\text{exp}}} ,\\\quad \text{for} \ {\lambda}_{ind}\in\ I_{ind}
% \end{array}
% \right\}.
% \end{aligned}
% \end{equation}
% where $\epsilon_{\DeltaT^{\text{exp}}}$ denotes the experimental uncertainty in measuring $\DeltaT^{\text{exp}}$.

As anticipated, the uncertainty in $\hat{\theta}_e$ becomes large when the NV and CuPc spectral features do not significantly overlap. However, at magnetic fields where the NV and CuPc spectra intersect, we obtain a precise estimate of $\theta_e$, as shown in Fig.~\ref{fig:orientation}. We note that, at a magnetic field of approximately 370~G, we observe two possible orientation angles, reflecting two possible orientations of the CuPc spectrum that give rise to the same spectral overlap. 
% As mentioned above, we can reconstruct the experiment data from the estimated parameters and theoretical model, shown in Fig.\ref{fig:T1_measurement}(c), where the errorbar is the fitting confidence interval. 
\subsection{CuPc hyperfine structure and Free radicals}
To uniquely identify CuPc as the cause for the change in $T_1$, we exploit its hyperfine fingerprint to distinguish it from other potential sources of spin noise. While we verified that no electron spins were present on the diamond surface before CuPc deposition, contamination could have occurred during deposition~\cite{sangtawesin2019origins,grinolds2014subnanometre,sushkov2014magnetic}. These electron spins typically exhibit properties similar to free electrons, characterized by a $g$-factor close to 2 and negligible hyperfine splitting. In addition, the CuPc layer could introduce other electronic spins. A radical electron due to oxidation of CuPc has been observed in electron paramagnetic resonance (EPR) experiments~\cite{atzori2016room,warner2013potential}. Photo-excitation induces the formation of a spin-triplet state in phthalocyanine, which however has a short lifetime (less than 400~ns at cryogenic temperatures) in the dark~\cite{barbon2001photoexcited}. 
%In CuPc, rather than the electron spin from $\text{Cu}^{2+}$ with strong hyperfine interaction with the nuclear spins, another potential electron spin source is the radical electron due to oxidation of CuPc, which have been observed in electron paramagnetic resonance (EPR) experiments~\cite{atzori2016room,warner2013potential}. Or other electron spin on the surface that could be enhanced  Previous research has demonstrated the formation of a spin-triplet state in phthalocyanine upon photo-excitation. Although this state may arise during photon illumination, it should not be considered a significant spin source due to its short lifetime (less than 400~ns at cryogenic temperatures) in the dark~\cite{barbon2001photoexcited}. 

To discriminate these potential spins from the $\text{Cu}^{2+}$ 
 electron spin  with its strong hyperfine interaction, we repeated the fitting procedure~\eqref{eq:fitting_value} and~\eqref{eq:fitting_error} assuming a spin bath composed solely of free electrons, with a spin density comparable to that of CuPc. As shown in Fig.~\ref{fig:T1_measurement}(c), % the discrepancy between the optimal fitting and experiment  remains robust regardless of the assumed correlation time for the free-electron bath, because 
 the free-electron spin bath model consistently yields spectral density values $S_e(\omega_{\text{NV}})$ too low to explain the observed reduction in $T_1$, even at magnetic fields around 370~G and 460~G, where free electron have small overlap with the NV. We note that the assumed
density of such free electron spins is already much higher than one would expect (for example, the  density of radical spins arising from oxidation processes is expected to be significantly lower, approximately 1\%, than the electron spin density of the CuPc layer.)
%, further diminishing their potential contribution. 

Thus our experiments not only unequivocally conclude that  CuPc electron spins dominate the NV depolarization but indirectly also reveal  the nuclear spin presence.
\subsection{Correlation time}
To understand the  mechanisms dominating the correlation time $\tau_e$ of the CuPc electron spins, we considered multiple possible decoherence contributions to the total decoherence rate, including  spin-lattice ($R_{s-l}$), electron-nuclear spin ($R_{e-n}$) and electron-electron ($R_{e-e}$) interactions. 
To evaluate the first two contributions, 
 we performed electron paramagnetic resonance (EPR) experiments on CuPc molecules diluted in a diamagnetic NiPc crystal at varying cryogenic temperature. Extrapolating these results to room temperature and assuming that these contributions do not depend on the CuPc dilution~\cite{ho2018spin,zhou2024phononic}, we predict that they would lead to a long $T_1\approx38$ns, significantly longer than the experimentally observed correlation time of $\sim 2.0$~ns. Details are shown in a separate work\cite{li25eprpaper}.
 %Because of the high similarity in nuclear spin environment, crystal structure and weight between CuPc and NiPc~\cite{robertson1935136}, and the dominate effect of crystal field within CuPc molecule instead of the inter molecule spin-lattice interaction~\cite{ho2018spin,zhou2024phononic}, these measurements allows us to infer the correlation time of pure CuPc due to combined effect from spin-lattice and electron-nuclear interaction for electron spin of a pure CuPc crystal at room temperature, which is approximately $38$~ns, significantly longer than the experimentally observed correlation time of $\sim 2.0$~ns.

This discrepancy indicates that electron-electron interactions ($R_{e-e}$) among densely packed CuPc electron spins in the pure CuPc crystal significantly enhance spin decoherence. To quantitatively evaluate this contribution, we numerically calculated the decoherence rate arising from rapid flip-flop interactions between electron spins localized on different CuPc molecules. The numerical results yield a decoherence rate consistent with our experimental estimation ($\tau_e|_{e\text{-}e} \in [1.3, 2.4]~\text{ns}$). Details of the simulations are shown in Appendix.~\ref{app:cupc_correlation_time}.

Combining these contributions, the total correlation rate of CuPc electron spins is given by:
\begin{equation}
    \frac{1}{\tau_e} = R_{s-l} + R_{e-n} + R_{e-e}.
\label{eq:cupc_correlation}
\end{equation}
Our analysis confirms that electron-electron interactions dominate the observed correlation time in pure CuPc crystals. We note that an independent measurement of the CuPc $\tau_e$ at room temp and 100\% density was not possible using bulk EPR methods, as the decay time is shorter than the probe capabilities.The close agreement between numerical calculations and experimental measurements demonstrates that our NV-based relaxometry method can locally probe spin dynamics on timescales inaccessible to conventional EPR spectroscopy.

\section{Estimation of the NV Center Depth}
\begin{figure}[htbp]
\centering
\includegraphics[width=0.8\linewidth]{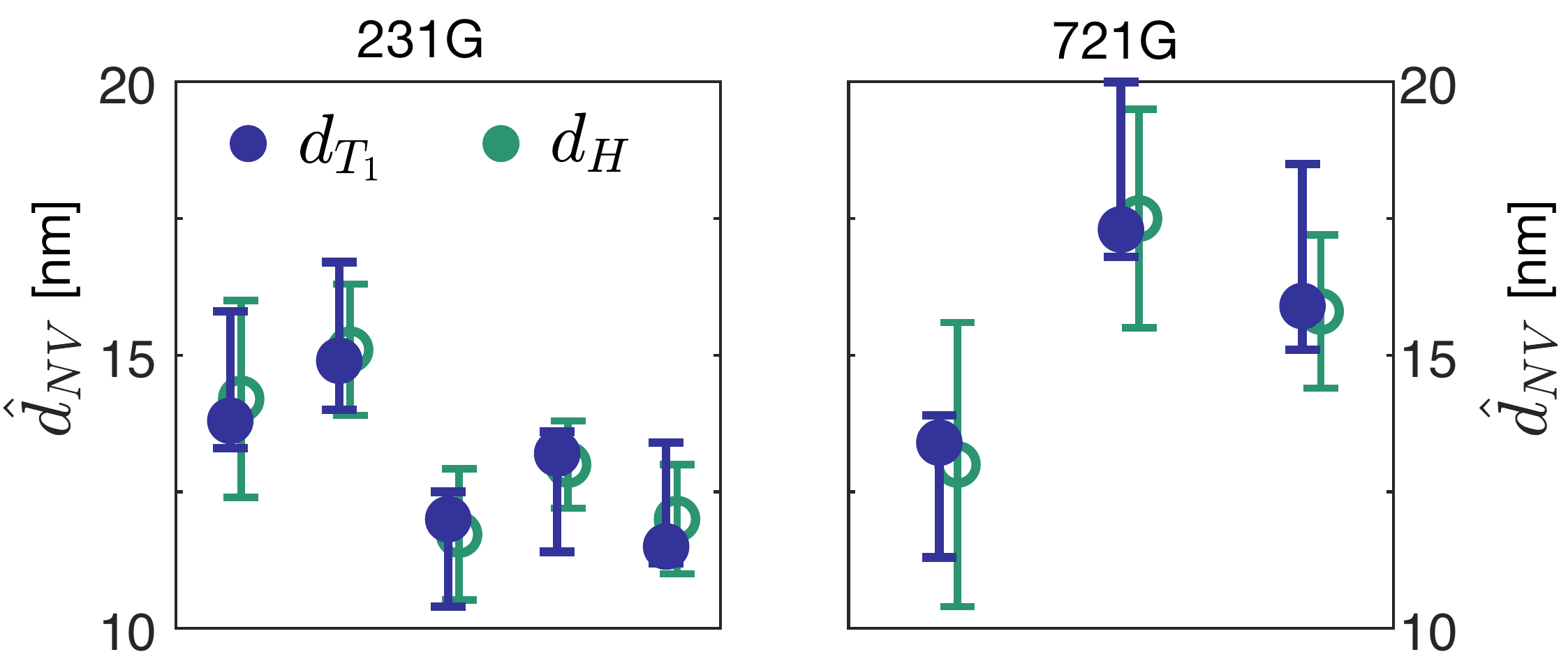}
\caption{NV center depth estimation. Data were collected at 231G and 721G. $d_{T_1}$ represents the depth extracted from the $T_1$ relaxometry measurements. $d_H$ denotes the depth measured after removing the CuPc layer using high-power laser illumination and exploiting the NV coupling to the hydrogen (H) in the objective oil~\cite{pham2016nmr}.  These depths are used  fixed parameters when estimating $\tau_e$ and $\theta_e$.
}
\label{fig:depth}
\end{figure}

\begin{table*}[t]
\caption{Comparison of different NV center depth measurement methods. Microwave-free: no microwave control is required throughout the experiment. Single calibration: different NV centers share the same experimental conditions; in our method, these conditions are the laser power and illumination time. Fixed magnetic field: the measurement can be performed without varying the magnetic field. Suitable for surface electron-free diamond: because surface electron spins degrade the coherence time of shallow NV centers, diamonds with surface treatments that remove surface spins are preferred for quantum applications.
}
\label{table:depth}

\begin{ruledtabular}
\begin{tabular}{lccc}
 & $T_1$ relaxometry (this work)
 & AFM scanning DEER~\cite{grinolds2014subnanometre}
 & NMR noise spectroscopy~\cite{pham2016nmr} \\
\hline
Precision             & nm       & sub-nm  & nm \\
Interaction           & Electron--electron spin & Electron--electron spin & Electron--nuclear spin \\
Protocol              & $T_1$ relaxation & Double electron--electron resonance (DEER) & Dynamical decoupling \\
Microwave-free        & \textbf{Yes} & No & No \\
Single calibration    & \textbf{Yes} & No & No \\
Fixed magnetic field  & \textbf{Yes} & No & \textbf{Yes} \\
No surface electrons  & \textbf{Yes} & No & \textbf{Yes} \\
\end{tabular}
\end{ruledtabular}
\end{table*}

% \begin{table*}[hpbt]
% \centering
% \begin{tabular}{>{\raggedright}m{1.2in}>{\centering\arraybackslash}m{1.8in}>{\centering\arraybackslash}m{1.8in}>{\centering\arraybackslash}m{1.8in}}
% \Xhline{1.2pt}
%  &   $T_1$ relaxation (this method) & AFM scanning DEER~\cite{grinolds2014subnanometre} & NMR noise spectroscopy ~\cite{pham2016nmr} \\
% \Xhline{1.2pt}
% Precision &   nm & sub-nm & nm \\
% \hline
% Interaction &   Electron-electron spin   & Electron-electron spin   & Electron-nuclear spin    \\
% \hline
%  Experiment
% & $T_1$ relaxation
% & Double electron-electron resonance (DEER) & Dynamical decoupling
% \\\hline
%  Microwave-free & \textbf{Yes} & No & No \\
%  Single calibration & \textbf{Yes} & No & No \\
%  Fixed magnetic field & \textbf{Yes} & No & \textbf{Yes} \\
% {No surface electrons} & \textbf{Yes} & No & \textbf{Yes} \\
% \Xhline{1.2pt}
% \end{tabular}
% \caption{Comparison of different NV center depth measurement methods. Microwave-free: no microwave control is required throughout the experiment. Single calibration: different NV centers share the same experimental conditions; in our method, these conditions are the laser power and illumination time. Fixed magnetic field: the measurement can be performed without varying the magnetic field. Suitable for surface electron-free diamond: because surface electron spins degrade the coherence time of shallow NV centers, diamonds with surface treatments that remove surface spins are preferred for quantum applications.
% }
% \label{table:depth}
% \end{table*}
While so far we assumed independent knowledge of the NV depth, $d_{NV}$ could have been extracted by fitting Eq.~\ref{eq:spec} and \ref{eq:T1_mech} to the data. This leads to  an alternative method to estimate the depth of shallow NV centers by exploiting their interaction with an external electron spin layer (in this case, a CuPc thin film). 
Making accurate depth characterization is essential  for quantum sensing  applications using shallow NV centers as it directly determines their coupling to external fields or qubits. Existing depth measurement techniques typically rely on  AFM scanning~\cite{grinolds2014subnanometre} or dynamical-decoupling (DD)~\cite{pham2016nmr} methods, which can achieve a precision of $\sim 1$~nm but require precise motion control at the nanometer level or high-fidelity microwave pulses. The
NV-$T_1$ relaxometry technique, on the other hand, is purely optical and compatible with standard confocal imaging setups. 

To demonstrate the robust estimate of the NV depth, we again focus on magnetic fields (231~G and 721~G) at which NV and CuPc transitions are sufficiently detuned, thus eliminating the $T_1$ dependence on the CuPc molecular orientation. We further assume that the electron spin ensemble correlation time $\tau_e$ can be independently estimated via numerical calculations of internal CuPc spin-spin interactions, as demonstrated in the preceding section, and that the electron spin layer can be removed by laser illumination. 
Employing the same nonlinear least-squares fitting procedure described above with $d_\textrm{NV}$ and $\theta_e$ as the fitting parameters, we extract the NV center depths, as shown in Fig.~\ref{fig:depth}. Excellent agreement is observed between our $T_1$-based depth estimate ($d_{T_1}$) and independent measurements based on detecting the protons   in the immersion oil surrounding the diamond sample after CuPc removal~\cite{pham2016nmr}, by applying  DD sequences matching the proton resonance frequency. 

The confidence intervals shown in Fig.~\ref{fig:depth}, which incorporate uncertainties in CuPc orientation and correlation time, are comparable to those obtained using nano-NMR-based depth characterization, highlighting the accuracy of our approach. Moreover, since $T_1$ measurements do not require individual calibration for each NV center, and the CuPc layer can be optically removed without damaging the diamond, our $T_1$-based depth estimation method offers a highly scalable alternative for NV center depth characterization. A comparison between the three methods are listed in Table~\ref{table:depth}.

\section{Conclusion and Outlook}

In this work, we demonstrated the ability to characterize the interaction between  shallow NV centers in diamond and the electron spins of a CuPc thin film using $T_1$ relaxometry. By applying our theoretical model and numerical fitting method, we extracted key properties of the CuPc spin bath, including its correlation time, local lattice orientation, and its contribution to NV depolarization. Our results confirm that electron-electron spin interactions are the dominant decoherence mechanism in CuPc at room temperature.  

Furthermore, we demonstrated a potential method for determining local crystal domain orientations in polycrystalline CuPc thin films, though further validation remains challenging due to the lack of established nanoscale structural characterization techniques .  Additionally, we introduced a novel approach for NV depth estimation using $T_1$ relaxometry, which does not require microwave pulses or prior knowledge of the CuPc lattice orientation. The extracted NV depths show strong agreement with conventional proton resonance measurements.

Looking ahead, our approach highlights the potential of NV centers as versatile quantum sensors for molecular spin systems and lays the foundation for further exploration of molecular qubits, spin bath engineering, and hybrid quantum materials at the nanoscale. The protocol could be extended to study other molecular spin systems and hybrid quantum materials. Beyond sensing, this research paves the way for leveraging such molecular systems for quantum logic operations, nuclear spin registers, and scalable entangled networks with fast, local control.

\ $Acknowledgements -$
This work has been supported by Honda Research Institute USA Inc.  
\appendix
\section{Autocorrelation Function of the CuPc Electron Spin Bath \label{app:auto_correlation_function}}

\subsection{Spin bath autocorrelation and spectral density}
\paragraph{CuPc-NV interaction}
The effective magnetic field from a CuPc electron spin at the NV center’s position arises from magnetic dipolar coupling:
\begin{equation}
    \vec{B} \cdot \vec{S}_e = \frac{\mu_0 \hbar \gamma_e}{4\pi} \left( \frac{3\vec{r}\vec{r} - \vec{r}^2 \mathbb{I}}{|\vec{r}|^5} \right) \cdot \vec{S}_e = \vec{D} \cdot \vec{S}_e,
\end{equation}
where $\vec{D}$ is the dipolar interaction tensor. Only the transverse ($x$- and $y$-) components of $\vec{B}$ contribute to the NV center spin relaxation time $T_1$.

For an ensemble of unpolarized CuPc spins, the transverse field autocorrelation is:
\begin{equation}
\begin{aligned}
    \sum_{\mu = x,y} \langle B_\mu(0) B_\mu(t) \rangle = \sum_{\mu,\nu,\nu'} \sum_{n,n'} D_{\mu\nu}^n D_{\mu\nu'}^{n'} \langle S_\nu^n(0) S_{\nu'}^{n'}(t) \rangle.
\end{aligned}
\label{eq:G4}
\end{equation}
Since the spins are in an unpolarized state, we can assume that different spins are uncorrelated, which also leads to preserving only the terms $\nu = \nu'$. Since all CuPc are identical, we have:    
\begin{equation}
\begin{aligned}   
\sum_{\mu = x,y}\langle B_{\mu}(0)B_{\mu}(t)\rangle=   \sum_{\mu} \sum_{\nu= x,y,z} \sum_{n}(D_{\mu\nu}^{n})^2\langle S_{\nu}(0)S_{\nu}(t) \rangle.
 \end{aligned}
 \label{eq:G1}
\end{equation}
Assuming the CuPc layer forms a uniform thin film of thickness $h$ and spin density $n_e$ on the diamond surface, and that the NV lies at depth $d_{\text{NV}}$, the sum of the dipolar coupling strength over all of the CuPc can be calculated as:
\begin{equation}
\begin{aligned}
    b_z^2 &= \int n_e \left( D_{xz}^2 + D_{yz}^2 \right) \mathrm{d}V = \frac{5}{16} b_0^2, \\
    b_\perp^2 &= \int n_e \left( D_{xx}^2 + D_{xy}^2 + D_{yx}^2 + D_{yy}^2 \right) \mathrm{d}V = \frac{11}{16} b_0^2, \\
    b_0^2 &= \left( \frac{\mu_0 \hbar \gamma_e}{4\pi} \right)^2 \frac{2\pi S_e(S_e+1)}{9} n_e \left( \frac{1}{d_{\text{NV}}^3} - \frac{1}{(d_{\text{NV}} + h)^3} \right),
\end{aligned}
\end{equation}
where $S_e(S_e+1) = 3/4$ is the total angular momentum for an electron spin-1/2. Here we separates the autocorrelation into longitudinal and transverse components corresponding to longitudinal and transverse spin autocorrelation:
\begin{equation}
\langle B_\perp(0) B_\perp(t) \rangle = b_\perp^2 \langle S_\perp(0) S_\perp(t) \rangle + b_z^2 \langle S_z(0) S_z(t) \rangle.
\label{eq:G3}
\end{equation}

To include multiple Cu isotopes (e.g., $^{63}\mathrm{Cu}$ and $^{65}\mathrm{Cu}$), each contributing distinct hyperfine structures and spin densities, we rewrite Eq.~\eqref{eq:G3} as:
\begin{equation}
\begin{aligned}
\langle B_\perp(0)& B_\perp(t) \rangle =\\
&\sum_{\kappa} \rho_\kappa \left[ b_\perp^2 \langle S_\perp(0) S_\perp(t) \rangle_\kappa + b_z^2 \langle S_z(0) S_z(t) \rangle_\kappa \right],
\end{aligned}
\label{eq:G2}
\end{equation}
where $\kappa$ indexes isotopes and $\rho_\kappa$ is the natural abundance.

\paragraph{CuPc electron spin states}
Each CuPc molecule experiences both Zeeman interaction and hyperfine couplings with its constituent nitrogen and copper nuclei. For $^{14}\mathrm{N}$ ($I = 1$), the hyperfine coupling constants are $A_{xx}^N = 57~\mathrm{MHz}$ and $A_{yy}^N = A_{zz}^N = 45~\mathrm{MHz}$~\cite{finazzo2006matrix}. Copper exists as two naturally abundant isotopes, $^{63}\mathrm{Cu}$ and $^{65}\mathrm{Cu}$, both with nuclear spin $I = \tfrac{3}{2}$. For $^{63}\mathrm{Cu}$ (69.15\% abundance), the electron hyperfine couplings are $A_{xx}^{\mathrm{Cu}} = A_{yy}^{\mathrm{Cu}} = -83~\mathrm{MHz}$ and $A_{zz}^{\mathrm{Cu}} = -648~\mathrm{MHz}$~\cite{finazzo2006matrix}.  The hyperfine constants for $^{65}\mathrm{Cu}$ (30.85\%) scale with its gyromagnetic ratio (${\gamma_{^{65}\mathrm{Cu}}}/{\gamma_{^{63}\mathrm{Cu}}} = 1.07$~\cite{stone2019table}).

For either isotope of copper, hyperfine interactions split the electron spin states into $2M$ ($M = 324$) hyperfine levels, with eigenstates $\ket{\psi_i}$ and eigen-frequencies $\omega_i$, for $i = 1, \dots, 2M$. Under the assumption of zero polarization at room temperature, each state is equally populated with a probability of $1/2M$. Transitions between states $\ket{\psi_i}$ and $\ket{\psi_j}$ occur at frequencies $\omega_{ij} = \omega_i - \omega_j$,  provided the transition matrix element $|\langle \psi_i | S_{\perp} | \psi_j \rangle|$ is nonzero. To account for both the transition strength and the uniform population, we define the  weight factor for each $i\leftrightarrow j$ transition:

\begin{equation}
\eta_{ij} = \frac{|\langle \psi_i | S_\perp | \psi_j \rangle|^2}{M}.
\end{equation}

Each $\ket{\psi_i} \leftrightarrow \ket{\psi_j}$ pair behaves as an effective spin-1/2 undergoing decoherence at rate $1/\tau_e$, modeled with the Lindblad master equation:
\begin{equation}
\begin{aligned}
    \frac{d\rho_{ij}}{dt} &= -i \omega_{ij} [\sigma_z, \rho_{ij}] \\
    &+ \frac{1}{\tau_e} \sum_{\mathcal{L}_k = \sigma_x, \sigma_y, \sigma_z} \left( \frac{1}{2} \mathcal{L}_k^\dagger \mathcal{L}_k \rho_{ij} + \frac{1}{2} \rho_{ij} \mathcal{L}_k^\dagger \mathcal{L}_k - \mathcal{L}_k \rho_{ij} \mathcal{L}_k^\dagger \right).
\end{aligned}
\end{equation}
This yields the correlation functions:
\begin{equation}
\begin{aligned}
    \langle S_z^{ij}(0) S_z^{ij}(t) \rangle &= e^{-|t|/\tau_e}, \\
    \langle S_\perp^{ij}(0) S_\perp^{ij}(t) \rangle &= e^{-|t|/\tau_e} \cos(\omega_{ij} t).
\end{aligned}
\end{equation}
The total bath correlation is then:
\begin{equation}
\langle S(0) S(t) \rangle = \sum_{i,j} \eta_{ij} \langle S^{ij}(0) S^{ij}(t) \rangle.
\end{equation}

\subsection{Power spectral density of the CuPc spin bath and NV longitudinal relaxation}
Combining the field fluctuation strength and spin dynamics, the autocorrelation function becomes:
\begin{equation}
\begin{aligned}
G_e(t) = \sum_{\kappa}\rho_{\kappa} \left\{ b_\perp^2 e^{-t/\tau_e} \sum_{i,j} \eta_{ij}^{(\kappa)} \cos(\omega_{ij}^{(\kappa)} t) + b_z^2 e^{-t/\tau_e} \right\},
\end{aligned}
\end{equation}
where $\kappa$ indexes isotopes and $\rho_\kappa$ is the natural abundance. The power spectral density $S_e(\omega)$ is obtained via Fourier transform:
\begin{widetext}
\begin{equation}
\begin{aligned}
S_e(\omega) = \sum_{\kappa} \rho_{\kappa}\left\{
\frac{5}{8} \cdot \frac{b_0^2 \tau_e}{\omega^2 \tau_e^2 + 1}
+ \frac{11}{16} \sum_{i,j} \eta_{ij}^{(\kappa)} \left[
\frac{b_0^2 \tau_e}{(\omega_{ij}^{(\kappa)} - \omega)^2 \tau_e^2 + 1}
+ \frac{b_0^2 \tau_e}{(\omega_{ij}^{(\kappa)} + \omega)^2 \tau_e^2 + 1}
\right] \right\}.
\end{aligned}
\end{equation}
\end{widetext}

Using Fermi's golden rule~\cite{machado2023quantum,rodriguez2018probing,choi2017depolarization,slichter2013principles,cohen1998atom}, the depolarization of NV center due to this fluctuating magnetic field is given by:
\begin{equation}
\begin{aligned}
     \frac{1}{T_1} = &\Gamma_{0\rightarrow1} + \Gamma_{1\rightarrow0}\\
 =&   \frac{1}{\hbar^2}\int_{-\infty}^{+\infty}dte^{i\omega_{NV} t}(H_{01}(0)H_{01}^{\dagger}(t)+H_{10}(0)H_{10}^{\dagger}(t)).
\end{aligned}
\label{eq:fermi's}
\end{equation}
The matrix element $H_{01/10}$ due to the fluctuating magnetic field is defined as:
\begin{equation}
\begin{aligned}
     H_{01} =&\hbar\gamma_e \langle0|B_xS_x+B_yS_y|1\rangle\
     = \hbar\gamma_e \frac{B_x+iB_y}{2}\langle0|S_{-}|1\rangle\\=&\hbar\gamma_e \frac{B_x+iB_y}{\sqrt{2}}.
\end{aligned}
\end{equation}
Here $S_{-} = S_x-iS_y$ is the ladder operator, and $\langle0|S_{-}|1\rangle=\sqrt{2}$
for spin-1. And therefore,  we obtain the $T_1$ of NV center as:
\begin{equation}
\begin{aligned}
        \frac{1}{T_1} &= \gamma_e^2\int_{-\infty}^{+\infty}dte^{i\omega_{NV} t}\left[B_x(0)B_x(t)+B_y(0)B_y(t)\right]\\
        &=\gamma_e^2\int_{-\infty}^{+\infty}dte^{i\omega_{NV} t}G_e(t)=\gamma_e^2S_e(\omega_{NV}).
\end{aligned}
\end{equation}
Here the autocorrelation function $G_e(t) =B_x(0)B_x(t)+B_y(0)B_y(t) $ is the same as the definition in Eqs.~\eqref{eq:G0} and~\eqref{eq:G4} 
\section{Correlation time estimation of CuPc electron spin \label{app:cupc_correlation_time}}
\paragraph{Spin-lattice interaction}
In the previous section, we described the decoherence of CuPc electron spins using a Lindblad In the previous section we considered a simple model for the decoherence of the CuPc states, described by a Lindblad master equation.
To evaluate the decoherence rate $\tau_e$ of CuPc electron spins, we consider that it arises from multiple sources, as discussed in the main text (Eq.~\eqref{eq:cupc_correlation}). The contribution from spin-lattice interactions, which is temperature-dependent, can be estimated using a diluted sample of CuPc in NiPc. NiPc has a highly similar crystal structure and molecular weight to CuPc. Additionally, in molecular systems, the spin-lattice interaction is predominantly determined by the crystal field inside each molecule~\cite{ho2018spin,zhou2024phononic}. Therefore, we can reasonably assume that the spin-lattice contribution to decoherence is independent of dilution.

We measured with ensemble ESR techniques~\cite{li25eprpaper} the depolarization time ($T_1$) of the CuPc electron spin at temperatures below 200~K and observed a strong temperature dependence, indicating that spin-lattice interactions dominate the decoherence processes in the diluted sample. By extrapolating the spin-lattice contribution to $T_1$ of CuPc electron spins to room temperature, we expect  an estimated $T_1 \approx 38~\text{ns}$.

\paragraph{Electron-electron interaction}
The pure CuPc thin film used in this study has a much higher electron spin density than the diluted sample. In this case, decoherence is dominated by electron-electron spin interactions between molecules, particularly through flip-flop processes. This contribution can be estimated similarly to how we calculate the decoherence effect of CuPc spins on the NV center.

For a given transition $\omega_{ij}$ in the $n$-th molecule, the electron spin experiences effective magnetic field fluctuations generated by all hyperfine transitions in neighboring CuPc molecules. According to Fermi’s golden rule, the depolarization rate is given by:
\begin{equation}
    \left(\frac{1}{\tau_e}\right)_{n,ij} = \frac{\gamma_e^2}{2} \sum_{m \ne n} S_{n,m}(\omega_{ij}),
\end{equation}
where $S_{n,m}(\omega_{ij})$ is the power spectral density of the magnetic noise generated by molecule $m$ at the position of molecule $n$.

Given the relative distance $r_{n,m}$ and orientation $\Theta_{n,m}$ between molecules $n$ and $m$, the spectral density is expressed as:
\begin{widetext}
\begin{equation}
\begin{aligned}
S_{n,m}(\omega_{ij}^{(\kappa)}) &= \left( \frac{\mu_0 \hbar \gamma_{e}}{4\pi} \right)^2 \frac{S_e(S_e+1)}{3r_{n,m}^6} \cdot
\left\{ \frac{9\sin^2(2\Theta_{n,m})}{2} \cdot \frac{\tau_e}{\omega_{ij}^2 \tau_e^2 + 1} \right. \\
&+ \left. \frac{5 - 6\cos(2\Theta_{n,m}) + 9\cos^2(2\Theta_{n,m})}{4} 
\sum_{\kappa'} \rho_{\kappa'} \sum_{i'j'} \eta_{i'j'}^{(\kappa')} \left[ \frac{\tau_e}{(\omega_{i'j'}^{(\kappa')} - \omega_{ij}^{(\kappa)})^2 \tau_e^2 + 1}
+ \frac{\tau_e}{(\omega_{i'j'}^{(\kappa')} + \omega_{ij}^{(\kappa)})^2 \tau_e^2 + 1} \right] \right\}.
\end{aligned}
\end{equation}
\end{widetext}

Here, $\kappa$ and $\kappa'$ index copper isotopes ($^{63}\mathrm{Cu}$ and $^{65}\mathrm{Cu}$), and $\rho_{\kappa’}$ is the natural abundance of isotope $\kappa’$. $\eta_{i’j’}^{(\kappa')}$ is the transition weight factor defined previously. Averaging over all transitions, molecules, and isotopes yields a self-consistent equation for $\tau_e$:
\begin{equation}
\frac{1}{\tau_e} = \left\langle \left( \frac{1}{\tau_e} \right)_{n,ij} \right\rangle = 
\gamma_e^2\sum_{\kappa} \rho_{\kappa} \sum_{ij} \eta_{ij}^{(\kappa)} 
\sum_{\substack{n,m \\ m \ne n}} S_{n,m}(\omega_{ij}^{(\kappa)}),
\label{eq:cupc_correlation_electron}
\end{equation}
which can be solved numerically, with lattice information in~\cite{luis2014molecular}.
The results show consistent predictions across the external magnetic fields used in our experiment, and exhibit variation when the orientation of the CuPc molecules relative to the external magnetic field is changed. The predicted electron-electron interaction limited correlation time spans the range $\tau_e|_{e\text{-}e} \in [1.3, 2.4]~\text{ns}$, which lies within the confidence interval of our experimentally extracted values.

An intuitive simplification of Eq.~\eqref{eq:cupc_correlation_electron} can be obtained by approximating the Lorentzian spectral densities as delta functions sharply peaked at the hyperfine transition frequencies. Under this first-order approximation:
\begin{equation}
\begin{aligned}
\frac{1}{\tau_e} 
&\approx \sqrt{\sum_{m \ne n} \sum_{\mu,\nu = x,y} (D_{\mu\nu}^{n,m})^2 \sum_{i'j',ij} 
\eta_{i'j'} \eta_{ij} \, \delta(\omega_{ij} - \omega_{i'j'})},
\end{aligned}
\end{equation}
where $D_{\mu\nu}^{n,m}$ is the dipolar interaction matrix elements. This expression captures flip-flop processes mediated by transverse dipolar couplings, which occur only when the energy splittings of two transitions exactly match. As a result, this approximation underestimates the total interaction strength, since in reality each spectral line has a finite linewidth. Numerically, this approach yields a correlation time of approximately $\sim 8~\text{ns}$.

Conversely, the interaction strength can be overestimated if the hyperfine structure is entirely ignored and all transitions are assumed to occur at the same frequency. In this extreme approximation, the correlation time becomes:
\begin{equation}
\frac{1}{\tau_e} 
\approx \sqrt{\sum_{m \ne n} \sum_{\mu,\nu = x,y} (D_{\mu\nu}^{n,m})^2} \sim 0.24~\text{ns}.
\end{equation}
This analysis highlights the necessity of incorporating a finite-width spectrum with hyperfine splitting in our model to accurately capture the spin bath dynamics.
Further numerical results and detailed analysis are provided in the Supplementary Materials.

\bibliography{CuPc_NV_ref}% Produces the bibliography via BibTeX.

\begin{thebibliography}{7}%
\makeatletter
\providecommand \@ifxundefined [1]{%
 \@ifx{#1\undefined}
}%
\providecommand \@ifnum [1]{%
 \ifnum #1\expandafter \@firstoftwo
 \else \expandafter \@secondoftwo
 \fi
}%
\providecommand \@ifx [1]{%
 \ifx #1\expandafter \@firstoftwo
 \else \expandafter \@secondoftwo
 \fi
}%
\providecommand \natexlab [1]{#1}%
\providecommand \enquote  [1]{``#1''}%
\providecommand \bibnamefont  [1]{#1}%
\providecommand \bibfnamefont [1]{#1}%
\providecommand \citenamefont [1]{#1}%
\providecommand \href@noop [0]{\@secondoftwo}%
\providecommand \href [0]{\begingroup \@sanitize@url \@href}%
\providecommand \@href[1]{\@@startlink{#1}\@@href}%
\providecommand \@@href[1]{\endgroup#1\@@endlink}%
\providecommand \@sanitize@url [0]{\catcode `\\12\catcode `\$12\catcode `\&12\catcode `\#12\catcode `\^12\catcode `\_12\catcode `\%12\relax}%
\providecommand \@@startlink[1]{}%
\providecommand \@@endlink[0]{}%
\providecommand \url  [0]{\begingroup\@sanitize@url \@url }%
\providecommand \@url [1]{\endgroup\@href {#1}{\urlprefix }}%
\providecommand \urlprefix  [0]{URL }%
\providecommand \Eprint [0]{\href }%
\providecommand \doibase [0]{http://dx.doi.org/}%
\providecommand \selectlanguage [0]{\@gobble}%
\providecommand \bibinfo  [0]{\@secondoftwo}%
\providecommand \bibfield  [0]{\@secondoftwo}%
\providecommand \translation [1]{[#1]}%
\providecommand \BibitemOpen [0]{}%
\providecommand \bibitemStop [0]{}%
\providecommand \bibitemNoStop [0]{.\EOS\space}%
\providecommand \EOS [0]{\spacefactor3000\relax}%
\providecommand \BibitemShut  [1]{\csname bibitem#1\endcsname}%
\let\auto@bib@innerbib\@empty
%</preamble>
\bibitem [{\citenamefont {Liu}\ \emph {et~al.}(2019)\citenamefont {Liu}, \citenamefont {Ajoy},\ and\ \citenamefont {Cappellaro}}]{liu2019nanoscale}%
  \BibitemOpen
  \bibfield  {author} {\bibinfo {author} {\bibfnamefont {Y.-X.}\ \bibnamefont {Liu}}, \bibinfo {author} {\bibfnamefont {A.}~\bibnamefont {Ajoy}}, \ and\ \bibinfo {author} {\bibfnamefont {P.}~\bibnamefont {Cappellaro}},\ }\href@noop {} {\bibfield  {journal} {\bibinfo  {journal} {Physical review letters}\ }\textbf {\bibinfo {volume} {122}},\ \bibinfo {pages} {100501} (\bibinfo {year} {2019})}\BibitemShut {NoStop}%
\bibitem [{\citenamefont {Sangtawesin}\ \emph {et~al.}(2019)\citenamefont {Sangtawesin}, \citenamefont {Dwyer}, \citenamefont {Srinivasan}, \citenamefont {Allred}, \citenamefont {Rodgers}, \citenamefont {De~Greve}, \citenamefont {Stacey}, \citenamefont {Dontschuk}, \citenamefont {O'Donnell}, \citenamefont {Hu} \emph {et~al.}}]{sangtawesin2019origins}%
  \BibitemOpen
  \bibfield  {author} {\bibinfo {author} {\bibfnamefont {S.}~\bibnamefont {Sangtawesin}}, \bibinfo {author} {\bibfnamefont {B.~L.}\ \bibnamefont {Dwyer}}, \bibinfo {author} {\bibfnamefont {S.}~\bibnamefont {Srinivasan}}, \bibinfo {author} {\bibfnamefont {J.~J.}\ \bibnamefont {Allred}}, \bibinfo {author} {\bibfnamefont {L.~V.}\ \bibnamefont {Rodgers}}, \bibinfo {author} {\bibfnamefont {K.}~\bibnamefont {De~Greve}}, \bibinfo {author} {\bibfnamefont {A.}~\bibnamefont {Stacey}}, \bibinfo {author} {\bibfnamefont {N.}~\bibnamefont {Dontschuk}}, \bibinfo {author} {\bibfnamefont {K.~M.}\ \bibnamefont {O'Donnell}}, \bibinfo {author} {\bibfnamefont {D.}~\bibnamefont {Hu}},  \emph {et~al.},\ }\href@noop {} {\bibfield  {journal} {\bibinfo  {journal} {Physical Review X}\ }\textbf {\bibinfo {volume} {9}},\ \bibinfo {pages} {031052} (\bibinfo {year} {2019})}\BibitemShut {NoStop}%
\bibitem [{\citenamefont {Sushkov}\ \emph {et~al.}(2014)\citenamefont {Sushkov}, \citenamefont {Lovchinsky}, \citenamefont {Chisholm}, \citenamefont {Walsworth}, \citenamefont {Park},\ and\ \citenamefont {Lukin}}]{sushkov2014magnetic}%
  \BibitemOpen
  \bibfield  {author} {\bibinfo {author} {\bibfnamefont {A.~O.}\ \bibnamefont {Sushkov}}, \bibinfo {author} {\bibfnamefont {I.}~\bibnamefont {Lovchinsky}}, \bibinfo {author} {\bibfnamefont {N.}~\bibnamefont {Chisholm}}, \bibinfo {author} {\bibfnamefont {R.~L.}\ \bibnamefont {Walsworth}}, \bibinfo {author} {\bibfnamefont {H.}~\bibnamefont {Park}}, \ and\ \bibinfo {author} {\bibfnamefont {M.~D.}\ \bibnamefont {Lukin}},\ }\href@noop {} {\bibfield  {journal} {\bibinfo  {journal} {Physical review letters}\ }\textbf {\bibinfo {volume} {113}},\ \bibinfo {pages} {197601} (\bibinfo {year} {2014})}\BibitemShut {NoStop}%
\bibitem [{\citenamefont {Shaibat}\ \emph {et~al.}(2010)\citenamefont {Shaibat}, \citenamefont {Casabianca}, \citenamefont {Siberio-P{\'e}rez}, \citenamefont {Matzger},\ and\ \citenamefont {Ishii}}]{shaibat2010distinguishing}%
  \BibitemOpen
  \bibfield  {author} {\bibinfo {author} {\bibfnamefont {M.~A.}\ \bibnamefont {Shaibat}}, \bibinfo {author} {\bibfnamefont {L.~B.}\ \bibnamefont {Casabianca}}, \bibinfo {author} {\bibfnamefont {D.~Y.}\ \bibnamefont {Siberio-P{\'e}rez}}, \bibinfo {author} {\bibfnamefont {A.~J.}\ \bibnamefont {Matzger}}, \ and\ \bibinfo {author} {\bibfnamefont {Y.}~\bibnamefont {Ishii}},\ }\href@noop {} {\bibfield  {journal} {\bibinfo  {journal} {The Journal of Physical Chemistry B}\ }\textbf {\bibinfo {volume} {114}},\ \bibinfo {pages} {4400} (\bibinfo {year} {2010})}\BibitemShut {NoStop}%
\bibitem [{\citenamefont {Pham}\ \emph {et~al.}(2016)\citenamefont {Pham}, \citenamefont {DeVience}, \citenamefont {Casola}, \citenamefont {Lovchinsky}, \citenamefont {Sushkov}, \citenamefont {Bersin}, \citenamefont {Lee}, \citenamefont {Urbach}, \citenamefont {Cappellaro}, \citenamefont {Park} \emph {et~al.}}]{pham2016nmr}%
  \BibitemOpen
  \bibfield  {author} {\bibinfo {author} {\bibfnamefont {L.~M.}\ \bibnamefont {Pham}}, \bibinfo {author} {\bibfnamefont {S.~J.}\ \bibnamefont {DeVience}}, \bibinfo {author} {\bibfnamefont {F.}~\bibnamefont {Casola}}, \bibinfo {author} {\bibfnamefont {I.}~\bibnamefont {Lovchinsky}}, \bibinfo {author} {\bibfnamefont {A.~O.}\ \bibnamefont {Sushkov}}, \bibinfo {author} {\bibfnamefont {E.}~\bibnamefont {Bersin}}, \bibinfo {author} {\bibfnamefont {J.}~\bibnamefont {Lee}}, \bibinfo {author} {\bibfnamefont {E.}~\bibnamefont {Urbach}}, \bibinfo {author} {\bibfnamefont {P.}~\bibnamefont {Cappellaro}}, \bibinfo {author} {\bibfnamefont {H.}~\bibnamefont {Park}},  \emph {et~al.},\ }\href@noop {} {\bibfield  {journal} {\bibinfo  {journal} {Physical Review B}\ }\textbf {\bibinfo {volume} {93}},\ \bibinfo {pages} {045425} (\bibinfo {year} {2016})}\BibitemShut {NoStop}%
\bibitem [{\citenamefont {Lunghi}(2022)}]{lunghi2022toward}%
  \BibitemOpen
  \bibfield  {author} {\bibinfo {author} {\bibfnamefont {A.}~\bibnamefont {Lunghi}},\ }\href@noop {} {\bibfield  {journal} {\bibinfo  {journal} {Science Advances}\ }\textbf {\bibinfo {volume} {8}},\ \bibinfo {pages} {eabn7880} (\bibinfo {year} {2022})}\BibitemShut {NoStop}%
\bibitem [{\citenamefont {Li}\ and\ \citenamefont {et~al.}()}]{li25eprpaper}%
  \BibitemOpen
  \bibfield  {author} {\bibinfo {author} {\bibfnamefont {B.}~\bibnamefont {Li}}\ and\ \bibinfo {author} {\bibnamefont {et~al.}},\ }\href@noop {} {\enquote {\bibinfo {title} {Exploring the mechanisms of transverse relaxation of copper(ii)-phthalocyanine spin qubits},}\ }\bibinfo {note} {Companion paper, submitted to arXiv (2025)}\BibitemShut {NoStop}%
\end{thebibliography}%


\begin{thebibliography}{50}%
\makeatletter
\providecommand \@ifxundefined [1]{%
 \@ifx{#1\undefined}
}%
\providecommand \@ifnum [1]{%
 \ifnum #1\expandafter \@firstoftwo
 \else \expandafter \@secondoftwo
 \fi
}%
\providecommand \@ifx [1]{%
 \ifx #1\expandafter \@firstoftwo
 \else \expandafter \@secondoftwo
 \fi
}%
\providecommand \natexlab [1]{#1}%
\providecommand \enquote  [1]{``#1''}%
\providecommand \bibnamefont  [1]{#1}%
\providecommand \bibfnamefont [1]{#1}%
\providecommand \citenamefont [1]{#1}%
\providecommand \href@noop [0]{\@secondoftwo}%
\providecommand \href [0]{\begingroup \@sanitize@url \@href}%
\providecommand \@href[1]{\@@startlink{#1}\@@href}%
\providecommand \@@href[1]{\endgroup#1\@@endlink}%
\providecommand \@sanitize@url [0]{\catcode `\\12\catcode `\$12\catcode `\&12\catcode `\#12\catcode `\^12\catcode `\_12\catcode `\%12\relax}%
\providecommand \@@startlink[1]{}%
\providecommand \@@endlink[0]{}%
\providecommand \url  [0]{\begingroup\@sanitize@url \@url }%
\providecommand \@url [1]{\endgroup\@href {#1}{\urlprefix }}%
\providecommand \urlprefix  [0]{URL }%
\providecommand \Eprint [0]{\href }%
\providecommand \doibase [0]{http://dx.doi.org/}%
\providecommand \selectlanguage [0]{\@gobble}%
\providecommand \bibinfo  [0]{\@secondoftwo}%
\providecommand \bibfield  [0]{\@secondoftwo}%
\providecommand \translation [1]{[#1]}%
\providecommand \BibitemOpen [0]{}%
\providecommand \bibitemStop [0]{}%
\providecommand \bibitemNoStop [0]{.\EOS\space}%
\providecommand \EOS [0]{\spacefactor3000\relax}%
\providecommand \BibitemShut  [1]{\csname bibitem#1\endcsname}%
\let\auto@bib@innerbib\@empty
%</preamble>
\bibitem [{\citenamefont {Steane}(1998)}]{steane1998quantum}%
  \BibitemOpen
  \bibfield  {author} {\bibinfo {author} {\bibfnamefont {A.}~\bibnamefont {Steane}},\ }\href@noop {} {\bibfield  {journal} {\bibinfo  {journal} {Reports on Progress in Physics}\ }\textbf {\bibinfo {volume} {61}},\ \bibinfo {pages} {117} (\bibinfo {year} {1998})}\BibitemShut {NoStop}%
\bibitem [{\citenamefont {Degen}\ \emph {et~al.}(2017)\citenamefont {Degen}, \citenamefont {Reinhard},\ and\ \citenamefont {Cappellaro}}]{degen2017quantum}%
  \BibitemOpen
  \bibfield  {author} {\bibinfo {author} {\bibfnamefont {C.~L.}\ \bibnamefont {Degen}}, \bibinfo {author} {\bibfnamefont {F.}~\bibnamefont {Reinhard}}, \ and\ \bibinfo {author} {\bibfnamefont {P.}~\bibnamefont {Cappellaro}},\ }\href@noop {} {\bibfield  {journal} {\bibinfo  {journal} {Reviews of modern physics}\ }\textbf {\bibinfo {volume} {89}},\ \bibinfo {pages} {035002} (\bibinfo {year} {2017})}\BibitemShut {NoStop}%
\bibitem [{\citenamefont {Cai}\ \emph {et~al.}(2013)\citenamefont {Cai}, \citenamefont {Retzker}, \citenamefont {Jelezko},\ and\ \citenamefont {Plenio}}]{cai2013large}%
  \BibitemOpen
  \bibfield  {author} {\bibinfo {author} {\bibfnamefont {J.}~\bibnamefont {Cai}}, \bibinfo {author} {\bibfnamefont {A.}~\bibnamefont {Retzker}}, \bibinfo {author} {\bibfnamefont {F.}~\bibnamefont {Jelezko}}, \ and\ \bibinfo {author} {\bibfnamefont {M.~B.}\ \bibnamefont {Plenio}},\ }\href@noop {} {\bibfield  {journal} {\bibinfo  {journal} {Nature Physics}\ }\textbf {\bibinfo {volume} {9}},\ \bibinfo {pages} {168} (\bibinfo {year} {2013})}\BibitemShut {NoStop}%
\bibitem [{\citenamefont {Tang}\ \emph {et~al.}(2023)\citenamefont {Tang}, \citenamefont {Li}, \citenamefont {Wang}, \citenamefont {Xu}, \citenamefont {Li}, \citenamefont {Barr}, \citenamefont {Cappellaro},\ and\ \citenamefont {Li}}]{tang2023communication}%
  \BibitemOpen
  \bibfield  {author} {\bibinfo {author} {\bibfnamefont {H.}~\bibnamefont {Tang}}, \bibinfo {author} {\bibfnamefont {B.}~\bibnamefont {Li}}, \bibinfo {author} {\bibfnamefont {G.}~\bibnamefont {Wang}}, \bibinfo {author} {\bibfnamefont {H.}~\bibnamefont {Xu}}, \bibinfo {author} {\bibfnamefont {C.}~\bibnamefont {Li}}, \bibinfo {author} {\bibfnamefont {A.}~\bibnamefont {Barr}}, \bibinfo {author} {\bibfnamefont {P.}~\bibnamefont {Cappellaro}}, \ and\ \bibinfo {author} {\bibfnamefont {J.}~\bibnamefont {Li}},\ }\href@noop {} {\bibfield  {journal} {\bibinfo  {journal} {Physical Review Letters}\ }\textbf {\bibinfo {volume} {130}},\ \bibinfo {pages} {150602} (\bibinfo {year} {2023})}\BibitemShut {NoStop}%
\bibitem [{\citenamefont {Li}\ \emph {et~al.}(2024)\citenamefont {Li}, \citenamefont {Li}, \citenamefont {Amer}, \citenamefont {Shaydulin}, \citenamefont {Chakrabarti}, \citenamefont {Wang}, \citenamefont {Xu}, \citenamefont {Tang}, \citenamefont {Schoch}, \citenamefont {Kumar} \emph {et~al.}}]{li2024blind}%
  \BibitemOpen
  \bibfield  {author} {\bibinfo {author} {\bibfnamefont {C.}~\bibnamefont {Li}}, \bibinfo {author} {\bibfnamefont {B.}~\bibnamefont {Li}}, \bibinfo {author} {\bibfnamefont {O.}~\bibnamefont {Amer}}, \bibinfo {author} {\bibfnamefont {R.}~\bibnamefont {Shaydulin}}, \bibinfo {author} {\bibfnamefont {S.}~\bibnamefont {Chakrabarti}}, \bibinfo {author} {\bibfnamefont {G.}~\bibnamefont {Wang}}, \bibinfo {author} {\bibfnamefont {H.}~\bibnamefont {Xu}}, \bibinfo {author} {\bibfnamefont {H.}~\bibnamefont {Tang}}, \bibinfo {author} {\bibfnamefont {I.}~\bibnamefont {Schoch}}, \bibinfo {author} {\bibfnamefont {N.}~\bibnamefont {Kumar}},  \emph {et~al.},\ }\href@noop {} {\bibfield  {journal} {\bibinfo  {journal} {Physical Review Letters}\ }\textbf {\bibinfo {volume} {133}},\ \bibinfo {pages} {120602} (\bibinfo {year} {2024})}\BibitemShut {NoStop}%
\bibitem [{\citenamefont {Abobeih}\ \emph {et~al.}(2022)\citenamefont {Abobeih}, \citenamefont {Wang}, \citenamefont {Randall}, \citenamefont {Loenen}, \citenamefont {Bradley}, \citenamefont {Markham}, \citenamefont {Twitchen}, \citenamefont {Terhal},\ and\ \citenamefont {Taminiau}}]{abobeih2022fault}%
  \BibitemOpen
  \bibfield  {author} {\bibinfo {author} {\bibfnamefont {M.~H.}\ \bibnamefont {Abobeih}}, \bibinfo {author} {\bibfnamefont {Y.}~\bibnamefont {Wang}}, \bibinfo {author} {\bibfnamefont {J.}~\bibnamefont {Randall}}, \bibinfo {author} {\bibfnamefont {S.}~\bibnamefont {Loenen}}, \bibinfo {author} {\bibfnamefont {C.~E.}\ \bibnamefont {Bradley}}, \bibinfo {author} {\bibfnamefont {M.}~\bibnamefont {Markham}}, \bibinfo {author} {\bibfnamefont {D.~J.}\ \bibnamefont {Twitchen}}, \bibinfo {author} {\bibfnamefont {B.~M.}\ \bibnamefont {Terhal}}, \ and\ \bibinfo {author} {\bibfnamefont {T.~H.}\ \bibnamefont {Taminiau}},\ }\href@noop {} {\bibfield  {journal} {\bibinfo  {journal} {Nature}\ }\textbf {\bibinfo {volume} {606}},\ \bibinfo {pages} {884} (\bibinfo {year} {2022})}\BibitemShut {NoStop}%
\bibitem [{\citenamefont {Godfrin}\ \emph {et~al.}(2017)\citenamefont {Godfrin}, \citenamefont {Ferhat}, \citenamefont {Ballou}, \citenamefont {Klyatskaya}, \citenamefont {Ruben}, \citenamefont {Wernsdorfer},\ and\ \citenamefont {Balestro}}]{godfrin2017operating}%
  \BibitemOpen
  \bibfield  {author} {\bibinfo {author} {\bibfnamefont {C.}~\bibnamefont {Godfrin}}, \bibinfo {author} {\bibfnamefont {A.}~\bibnamefont {Ferhat}}, \bibinfo {author} {\bibfnamefont {R.}~\bibnamefont {Ballou}}, \bibinfo {author} {\bibfnamefont {S.}~\bibnamefont {Klyatskaya}}, \bibinfo {author} {\bibfnamefont {M.}~\bibnamefont {Ruben}}, \bibinfo {author} {\bibfnamefont {W.}~\bibnamefont {Wernsdorfer}}, \ and\ \bibinfo {author} {\bibfnamefont {F.}~\bibnamefont {Balestro}},\ }\href@noop {} {\bibfield  {journal} {\bibinfo  {journal} {Physical review letters}\ }\textbf {\bibinfo {volume} {119}},\ \bibinfo {pages} {187702} (\bibinfo {year} {2017})}\BibitemShut {NoStop}%
\bibitem [{\citenamefont {Ruskuc}\ \emph {et~al.}(2022)\citenamefont {Ruskuc}, \citenamefont {Wu}, \citenamefont {Rochman}, \citenamefont {Choi},\ and\ \citenamefont {Faraon}}]{ruskuc2022nuclear}%
  \BibitemOpen
  \bibfield  {author} {\bibinfo {author} {\bibfnamefont {A.}~\bibnamefont {Ruskuc}}, \bibinfo {author} {\bibfnamefont {C.-J.}\ \bibnamefont {Wu}}, \bibinfo {author} {\bibfnamefont {J.}~\bibnamefont {Rochman}}, \bibinfo {author} {\bibfnamefont {J.}~\bibnamefont {Choi}}, \ and\ \bibinfo {author} {\bibfnamefont {A.}~\bibnamefont {Faraon}},\ }\href@noop {} {\bibfield  {journal} {\bibinfo  {journal} {Nature}\ }\textbf {\bibinfo {volume} {602}},\ \bibinfo {pages} {408} (\bibinfo {year} {2022})}\BibitemShut {NoStop}%
\bibitem [{\citenamefont {Lim}\ \emph {et~al.}(2025)\citenamefont {Lim}, \citenamefont {Vaganov}, \citenamefont {Liu},\ and\ \citenamefont {Ardavan}}]{lim2025demonstrating}%
  \BibitemOpen
  \bibfield  {author} {\bibinfo {author} {\bibfnamefont {S.}~\bibnamefont {Lim}}, \bibinfo {author} {\bibfnamefont {M.~V.}\ \bibnamefont {Vaganov}}, \bibinfo {author} {\bibfnamefont {J.}~\bibnamefont {Liu}}, \ and\ \bibinfo {author} {\bibfnamefont {A.}~\bibnamefont {Ardavan}},\ }\href@noop {} {\bibfield  {journal} {\bibinfo  {journal} {Physical Review Letters}\ }\textbf {\bibinfo {volume} {134}},\ \bibinfo {pages} {070603} (\bibinfo {year} {2025})}\BibitemShut {NoStop}%
\bibitem [{\citenamefont {Qiu}\ \emph {et~al.}(2021)\citenamefont {Qiu}, \citenamefont {Vool}, \citenamefont {Hamo},\ and\ \citenamefont {Yacoby}}]{qiu2021nuclear}%
  \BibitemOpen
  \bibfield  {author} {\bibinfo {author} {\bibfnamefont {Z.}~\bibnamefont {Qiu}}, \bibinfo {author} {\bibfnamefont {U.}~\bibnamefont {Vool}}, \bibinfo {author} {\bibfnamefont {A.}~\bibnamefont {Hamo}}, \ and\ \bibinfo {author} {\bibfnamefont {A.}~\bibnamefont {Yacoby}},\ }\href@noop {} {\bibfield  {journal} {\bibinfo  {journal} {npj Quantum Information}\ }\textbf {\bibinfo {volume} {7}},\ \bibinfo {pages} {39} (\bibinfo {year} {2021})}\BibitemShut {NoStop}%
\bibitem [{\citenamefont {Wang}\ \emph {et~al.}(2024{\natexlab{a}})\citenamefont {Wang}, \citenamefont {Nguyen},\ and\ \citenamefont {Cappellaro}}]{wang2024hyperfine}%
  \BibitemOpen
  \bibfield  {author} {\bibinfo {author} {\bibfnamefont {G.}~\bibnamefont {Wang}}, \bibinfo {author} {\bibfnamefont {M.-T.}\ \bibnamefont {Nguyen}}, \ and\ \bibinfo {author} {\bibfnamefont {P.}~\bibnamefont {Cappellaro}},\ }\href@noop {} {\bibfield  {journal} {\bibinfo  {journal} {Physical Review Letters}\ }\textbf {\bibinfo {volume} {133}},\ \bibinfo {pages} {150801} (\bibinfo {year} {2024}{\natexlab{a}})}\BibitemShut {NoStop}%
\bibitem [{\citenamefont {Atzori}\ \emph {et~al.}(2016)\citenamefont {Atzori}, \citenamefont {Tesi}, \citenamefont {Morra}, \citenamefont {Chiesa}, \citenamefont {Sorace},\ and\ \citenamefont {Sessoli}}]{atzori2016room}%
  \BibitemOpen
  \bibfield  {author} {\bibinfo {author} {\bibfnamefont {M.}~\bibnamefont {Atzori}}, \bibinfo {author} {\bibfnamefont {L.}~\bibnamefont {Tesi}}, \bibinfo {author} {\bibfnamefont {E.}~\bibnamefont {Morra}}, \bibinfo {author} {\bibfnamefont {M.}~\bibnamefont {Chiesa}}, \bibinfo {author} {\bibfnamefont {L.}~\bibnamefont {Sorace}}, \ and\ \bibinfo {author} {\bibfnamefont {R.}~\bibnamefont {Sessoli}},\ }\href@noop {} {\bibfield  {journal} {\bibinfo  {journal} {Journal of the American Chemical Society}\ }\textbf {\bibinfo {volume} {138}},\ \bibinfo {pages} {2154} (\bibinfo {year} {2016})}\BibitemShut {NoStop}%
\bibitem [{\citenamefont {Mullin}\ \emph {et~al.}(2024)\citenamefont {Mullin}, \citenamefont {Johnson}, \citenamefont {Freedman},\ and\ \citenamefont {Rondinelli}}]{mullin2024systems}%
  \BibitemOpen
  \bibfield  {author} {\bibinfo {author} {\bibfnamefont {K.~R.}\ \bibnamefont {Mullin}}, \bibinfo {author} {\bibfnamefont {D.}~\bibnamefont {Johnson}}, \bibinfo {author} {\bibfnamefont {D.~E.}\ \bibnamefont {Freedman}}, \ and\ \bibinfo {author} {\bibfnamefont {J.~M.}\ \bibnamefont {Rondinelli}},\ }\href@noop {} {\bibfield  {journal} {\bibinfo  {journal} {Dalton Transactions}\ }\textbf {\bibinfo {volume} {53}},\ \bibinfo {pages} {16585} (\bibinfo {year} {2024})}\BibitemShut {NoStop}%
\bibitem [{\citenamefont {Warner}\ \emph {et~al.}(2013)\citenamefont {Warner}, \citenamefont {Din}, \citenamefont {Tupitsyn}, \citenamefont {Morley}, \citenamefont {Stoneham}, \citenamefont {Gardener}, \citenamefont {Wu}, \citenamefont {Fisher}, \citenamefont {Heutz}, \citenamefont {Kay} \emph {et~al.}}]{warner2013potential}%
  \BibitemOpen
  \bibfield  {author} {\bibinfo {author} {\bibfnamefont {M.}~\bibnamefont {Warner}}, \bibinfo {author} {\bibfnamefont {S.}~\bibnamefont {Din}}, \bibinfo {author} {\bibfnamefont {I.~S.}\ \bibnamefont {Tupitsyn}}, \bibinfo {author} {\bibfnamefont {G.~W.}\ \bibnamefont {Morley}}, \bibinfo {author} {\bibfnamefont {A.~M.}\ \bibnamefont {Stoneham}}, \bibinfo {author} {\bibfnamefont {J.~A.}\ \bibnamefont {Gardener}}, \bibinfo {author} {\bibfnamefont {Z.}~\bibnamefont {Wu}}, \bibinfo {author} {\bibfnamefont {A.~J.}\ \bibnamefont {Fisher}}, \bibinfo {author} {\bibfnamefont {S.}~\bibnamefont {Heutz}}, \bibinfo {author} {\bibfnamefont {C.~W.}\ \bibnamefont {Kay}},  \emph {et~al.},\ }\href@noop {} {\bibfield  {journal} {\bibinfo  {journal} {Nature}\ }\textbf {\bibinfo {volume} {503}},\ \bibinfo {pages} {504} (\bibinfo {year} {2013})}\BibitemShut {NoStop}%
\bibitem [{\citenamefont {Wedge}\ \emph {et~al.}(2012)\citenamefont {Wedge}, \citenamefont {Timco}, \citenamefont {Spielberg}, \citenamefont {George}, \citenamefont {Tuna}, \citenamefont {Rigby}, \citenamefont {McInnes}, \citenamefont {Winpenny}, \citenamefont {Blundell},\ and\ \citenamefont {Ardavan}}]{wedge2012chemical}%
  \BibitemOpen
  \bibfield  {author} {\bibinfo {author} {\bibfnamefont {C.~J.}\ \bibnamefont {Wedge}}, \bibinfo {author} {\bibfnamefont {G.}~\bibnamefont {Timco}}, \bibinfo {author} {\bibfnamefont {E.}~\bibnamefont {Spielberg}}, \bibinfo {author} {\bibfnamefont {R.}~\bibnamefont {George}}, \bibinfo {author} {\bibfnamefont {F.}~\bibnamefont {Tuna}}, \bibinfo {author} {\bibfnamefont {S.}~\bibnamefont {Rigby}}, \bibinfo {author} {\bibfnamefont {E.}~\bibnamefont {McInnes}}, \bibinfo {author} {\bibfnamefont {R.}~\bibnamefont {Winpenny}}, \bibinfo {author} {\bibfnamefont {S.}~\bibnamefont {Blundell}}, \ and\ \bibinfo {author} {\bibfnamefont {A.}~\bibnamefont {Ardavan}},\ }\href@noop {} {\bibfield  {journal} {\bibinfo  {journal} {Physical Review Letters}\ }\textbf {\bibinfo {volume} {108}},\ \bibinfo {pages} {107204} (\bibinfo {year} {2012})}\BibitemShut {NoStop}%
\bibitem [{\citenamefont {Lavroff}\ \emph {et~al.}(2021)\citenamefont {Lavroff}, \citenamefont {Pennington}, \citenamefont {Hua}, \citenamefont {Li}, \citenamefont {Williams},\ and\ \citenamefont {Alexandrova}}]{lavroff2021recent}%
  \BibitemOpen
  \bibfield  {author} {\bibinfo {author} {\bibfnamefont {R.~H.}\ \bibnamefont {Lavroff}}, \bibinfo {author} {\bibfnamefont {D.~L.}\ \bibnamefont {Pennington}}, \bibinfo {author} {\bibfnamefont {A.~S.}\ \bibnamefont {Hua}}, \bibinfo {author} {\bibfnamefont {B.~Y.}\ \bibnamefont {Li}}, \bibinfo {author} {\bibfnamefont {J.~A.}\ \bibnamefont {Williams}}, \ and\ \bibinfo {author} {\bibfnamefont {A.~N.}\ \bibnamefont {Alexandrova}},\ }\href@noop {} {\enquote {\bibinfo {title} {Recent innovations in solid-state and molecular qubits for quantum information applications},}\ } (\bibinfo {year} {2021})\BibitemShut {NoStop}%
\bibitem [{\citenamefont {Gaita-Ari{\~n}o}\ \emph {et~al.}(2019)\citenamefont {Gaita-Ari{\~n}o}, \citenamefont {Luis}, \citenamefont {Hill},\ and\ \citenamefont {Coronado}}]{gaita2019molecular}%
  \BibitemOpen
  \bibfield  {author} {\bibinfo {author} {\bibfnamefont {A.}~\bibnamefont {Gaita-Ari{\~n}o}}, \bibinfo {author} {\bibfnamefont {F.}~\bibnamefont {Luis}}, \bibinfo {author} {\bibfnamefont {S.}~\bibnamefont {Hill}}, \ and\ \bibinfo {author} {\bibfnamefont {E.}~\bibnamefont {Coronado}},\ }\href@noop {} {\bibfield  {journal} {\bibinfo  {journal} {Nature chemistry}\ }\textbf {\bibinfo {volume} {11}},\ \bibinfo {pages} {301} (\bibinfo {year} {2019})}\BibitemShut {NoStop}%
\bibitem [{\citenamefont {Bayliss}\ \emph {et~al.}(2020)\citenamefont {Bayliss}, \citenamefont {Laorenza}, \citenamefont {Mintun}, \citenamefont {Kovos}, \citenamefont {Freedman},\ and\ \citenamefont {Awschalom}}]{bayliss2020optically}%
  \BibitemOpen
  \bibfield  {author} {\bibinfo {author} {\bibfnamefont {S.}~\bibnamefont {Bayliss}}, \bibinfo {author} {\bibfnamefont {D.}~\bibnamefont {Laorenza}}, \bibinfo {author} {\bibfnamefont {P.}~\bibnamefont {Mintun}}, \bibinfo {author} {\bibfnamefont {B.}~\bibnamefont {Kovos}}, \bibinfo {author} {\bibfnamefont {D.~E.}\ \bibnamefont {Freedman}}, \ and\ \bibinfo {author} {\bibfnamefont {D.}~\bibnamefont {Awschalom}},\ }\href@noop {} {\bibfield  {journal} {\bibinfo  {journal} {Science}\ }\textbf {\bibinfo {volume} {370}},\ \bibinfo {pages} {1309} (\bibinfo {year} {2020})}\BibitemShut {NoStop}%
\bibitem [{\citenamefont {Doherty}\ \emph {et~al.}(2013)\citenamefont {Doherty}, \citenamefont {Manson}, \citenamefont {Delaney}, \citenamefont {Jelezko}, \citenamefont {Wrachtrup},\ and\ \citenamefont {Hollenberg}}]{doherty2013nitrogen}%
  \BibitemOpen
  \bibfield  {author} {\bibinfo {author} {\bibfnamefont {M.~W.}\ \bibnamefont {Doherty}}, \bibinfo {author} {\bibfnamefont {N.~B.}\ \bibnamefont {Manson}}, \bibinfo {author} {\bibfnamefont {P.}~\bibnamefont {Delaney}}, \bibinfo {author} {\bibfnamefont {F.}~\bibnamefont {Jelezko}}, \bibinfo {author} {\bibfnamefont {J.}~\bibnamefont {Wrachtrup}}, \ and\ \bibinfo {author} {\bibfnamefont {L.~C.}\ \bibnamefont {Hollenberg}},\ }\href@noop {} {\bibfield  {journal} {\bibinfo  {journal} {Physics Reports}\ }\textbf {\bibinfo {volume} {528}},\ \bibinfo {pages} {1} (\bibinfo {year} {2013})}\BibitemShut {NoStop}%
\bibitem [{\citenamefont {Pham}\ \emph {et~al.}(2016)\citenamefont {Pham}, \citenamefont {DeVience}, \citenamefont {Casola}, \citenamefont {Lovchinsky}, \citenamefont {Sushkov}, \citenamefont {Bersin}, \citenamefont {Lee}, \citenamefont {Urbach}, \citenamefont {Cappellaro}, \citenamefont {Park} \emph {et~al.}}]{pham2016nmr}%
  \BibitemOpen
  \bibfield  {author} {\bibinfo {author} {\bibfnamefont {L.~M.}\ \bibnamefont {Pham}}, \bibinfo {author} {\bibfnamefont {S.~J.}\ \bibnamefont {DeVience}}, \bibinfo {author} {\bibfnamefont {F.}~\bibnamefont {Casola}}, \bibinfo {author} {\bibfnamefont {I.}~\bibnamefont {Lovchinsky}}, \bibinfo {author} {\bibfnamefont {A.~O.}\ \bibnamefont {Sushkov}}, \bibinfo {author} {\bibfnamefont {E.}~\bibnamefont {Bersin}}, \bibinfo {author} {\bibfnamefont {J.}~\bibnamefont {Lee}}, \bibinfo {author} {\bibfnamefont {E.}~\bibnamefont {Urbach}}, \bibinfo {author} {\bibfnamefont {P.}~\bibnamefont {Cappellaro}}, \bibinfo {author} {\bibfnamefont {H.}~\bibnamefont {Park}},  \emph {et~al.},\ }\href@noop {} {\bibfield  {journal} {\bibinfo  {journal} {Physical Review B}\ }\textbf {\bibinfo {volume} {93}},\ \bibinfo {pages} {045425} (\bibinfo {year} {2016})}\BibitemShut {NoStop}%
\bibitem [{\citenamefont {Sangtawesin}\ \emph {et~al.}(2019)\citenamefont {Sangtawesin}, \citenamefont {Dwyer}, \citenamefont {Srinivasan}, \citenamefont {Allred}, \citenamefont {Rodgers}, \citenamefont {De~Greve}, \citenamefont {Stacey}, \citenamefont {Dontschuk}, \citenamefont {O'Donnell}, \citenamefont {Hu} \emph {et~al.}}]{sangtawesin2019origins}%
  \BibitemOpen
  \bibfield  {author} {\bibinfo {author} {\bibfnamefont {S.}~\bibnamefont {Sangtawesin}}, \bibinfo {author} {\bibfnamefont {B.~L.}\ \bibnamefont {Dwyer}}, \bibinfo {author} {\bibfnamefont {S.}~\bibnamefont {Srinivasan}}, \bibinfo {author} {\bibfnamefont {J.~J.}\ \bibnamefont {Allred}}, \bibinfo {author} {\bibfnamefont {L.~V.}\ \bibnamefont {Rodgers}}, \bibinfo {author} {\bibfnamefont {K.}~\bibnamefont {De~Greve}}, \bibinfo {author} {\bibfnamefont {A.}~\bibnamefont {Stacey}}, \bibinfo {author} {\bibfnamefont {N.}~\bibnamefont {Dontschuk}}, \bibinfo {author} {\bibfnamefont {K.~M.}\ \bibnamefont {O'Donnell}}, \bibinfo {author} {\bibfnamefont {D.}~\bibnamefont {Hu}},  \emph {et~al.},\ }\href@noop {} {\bibfield  {journal} {\bibinfo  {journal} {Physical Review X}\ }\textbf {\bibinfo {volume} {9}},\ \bibinfo {pages} {031052} (\bibinfo {year} {2019})}\BibitemShut {NoStop}%
\bibitem [{\citenamefont {Sun}\ and\ \citenamefont {Cappellaro}(2022)}]{sun2022self}%
  \BibitemOpen
  \bibfield  {author} {\bibinfo {author} {\bibfnamefont {W.~K.~C.}\ \bibnamefont {Sun}}\ and\ \bibinfo {author} {\bibfnamefont {P.}~\bibnamefont {Cappellaro}},\ }\href@noop {} {\bibfield  {journal} {\bibinfo  {journal} {Physical Review B}\ }\textbf {\bibinfo {volume} {106}},\ \bibinfo {pages} {155413} (\bibinfo {year} {2022})}\BibitemShut {NoStop}%
\bibitem [{\citenamefont {Wang}\ \emph {et~al.}(2024{\natexlab{b}})\citenamefont {Wang}, \citenamefont {Zhu}, \citenamefont {Li}, \citenamefont {Li}, \citenamefont {Viola}, \citenamefont {Cooper},\ and\ \citenamefont {Cappellaro}}]{wang2024digital}%
  \BibitemOpen
  \bibfield  {author} {\bibinfo {author} {\bibfnamefont {G.}~\bibnamefont {Wang}}, \bibinfo {author} {\bibfnamefont {Y.}~\bibnamefont {Zhu}}, \bibinfo {author} {\bibfnamefont {B.}~\bibnamefont {Li}}, \bibinfo {author} {\bibfnamefont {C.}~\bibnamefont {Li}}, \bibinfo {author} {\bibfnamefont {L.}~\bibnamefont {Viola}}, \bibinfo {author} {\bibfnamefont {A.}~\bibnamefont {Cooper}}, \ and\ \bibinfo {author} {\bibfnamefont {P.}~\bibnamefont {Cappellaro}},\ }\href@noop {} {\bibfield  {journal} {\bibinfo  {journal} {Quantum Science and Technology}\ }\textbf {\bibinfo {volume} {9}},\ \bibinfo {pages} {035006} (\bibinfo {year} {2024}{\natexlab{b}})}\BibitemShut {NoStop}%
\bibitem [{\citenamefont {Sushkov}\ \emph {et~al.}(2014)\citenamefont {Sushkov}, \citenamefont {Lovchinsky}, \citenamefont {Chisholm}, \citenamefont {Walsworth}, \citenamefont {Park},\ and\ \citenamefont {Lukin}}]{sushkov2014magnetic}%
  \BibitemOpen
  \bibfield  {author} {\bibinfo {author} {\bibfnamefont {A.~O.}\ \bibnamefont {Sushkov}}, \bibinfo {author} {\bibfnamefont {I.}~\bibnamefont {Lovchinsky}}, \bibinfo {author} {\bibfnamefont {N.}~\bibnamefont {Chisholm}}, \bibinfo {author} {\bibfnamefont {R.~L.}\ \bibnamefont {Walsworth}}, \bibinfo {author} {\bibfnamefont {H.}~\bibnamefont {Park}}, \ and\ \bibinfo {author} {\bibfnamefont {M.~D.}\ \bibnamefont {Lukin}},\ }\href@noop {} {\bibfield  {journal} {\bibinfo  {journal} {Physical review letters}\ }\textbf {\bibinfo {volume} {113}},\ \bibinfo {pages} {197601} (\bibinfo {year} {2014})}\BibitemShut {NoStop}%
\bibitem [{\citenamefont {Cooper}\ \emph {et~al.}(2020)\citenamefont {Cooper}, \citenamefont {Sun}, \citenamefont {Jaskula},\ and\ \citenamefont {Cappellaro}}]{cooper2020identification}%
  \BibitemOpen
  \bibfield  {author} {\bibinfo {author} {\bibfnamefont {A.}~\bibnamefont {Cooper}}, \bibinfo {author} {\bibfnamefont {W.~K.~C.}\ \bibnamefont {Sun}}, \bibinfo {author} {\bibfnamefont {J.-C.}\ \bibnamefont {Jaskula}}, \ and\ \bibinfo {author} {\bibfnamefont {P.}~\bibnamefont {Cappellaro}},\ }\href@noop {} {\bibfield  {journal} {\bibinfo  {journal} {Physical Review Letters}\ }\textbf {\bibinfo {volume} {124}},\ \bibinfo {pages} {083602} (\bibinfo {year} {2020})}\BibitemShut {NoStop}%
\bibitem [{\citenamefont {Knowles}\ \emph {et~al.}(2016)\citenamefont {Knowles}, \citenamefont {Kara},\ and\ \citenamefont {Atat{\"u}re}}]{knowles2016demonstration}%
  \BibitemOpen
  \bibfield  {author} {\bibinfo {author} {\bibfnamefont {H.~S.}\ \bibnamefont {Knowles}}, \bibinfo {author} {\bibfnamefont {D.~M.}\ \bibnamefont {Kara}}, \ and\ \bibinfo {author} {\bibfnamefont {M.}~\bibnamefont {Atat{\"u}re}},\ }\href@noop {} {\bibfield  {journal} {\bibinfo  {journal} {Physical Review Letters}\ }\textbf {\bibinfo {volume} {117}},\ \bibinfo {pages} {100802} (\bibinfo {year} {2016})}\BibitemShut {NoStop}%
\bibitem [{\citenamefont {Rosenfeld}\ \emph {et~al.}(2018)\citenamefont {Rosenfeld}, \citenamefont {Pham}, \citenamefont {Lukin},\ and\ \citenamefont {Walsworth}}]{rosenfeld2018sensing}%
  \BibitemOpen
  \bibfield  {author} {\bibinfo {author} {\bibfnamefont {E.~L.}\ \bibnamefont {Rosenfeld}}, \bibinfo {author} {\bibfnamefont {L.~M.}\ \bibnamefont {Pham}}, \bibinfo {author} {\bibfnamefont {M.~D.}\ \bibnamefont {Lukin}}, \ and\ \bibinfo {author} {\bibfnamefont {R.~L.}\ \bibnamefont {Walsworth}},\ }\href@noop {} {\bibfield  {journal} {\bibinfo  {journal} {Physical review letters}\ }\textbf {\bibinfo {volume} {120}},\ \bibinfo {pages} {243604} (\bibinfo {year} {2018})}\BibitemShut {NoStop}%
\bibitem [{\citenamefont {Degen}\ \emph {et~al.}(2021)\citenamefont {Degen}, \citenamefont {Loenen}, \citenamefont {Bartling}, \citenamefont {Bradley}, \citenamefont {Meinsma}, \citenamefont {Markham}, \citenamefont {Twitchen},\ and\ \citenamefont {Taminiau}}]{degen2021entanglement}%
  \BibitemOpen
  \bibfield  {author} {\bibinfo {author} {\bibfnamefont {M.}~\bibnamefont {Degen}}, \bibinfo {author} {\bibfnamefont {S.}~\bibnamefont {Loenen}}, \bibinfo {author} {\bibfnamefont {H.}~\bibnamefont {Bartling}}, \bibinfo {author} {\bibfnamefont {C.}~\bibnamefont {Bradley}}, \bibinfo {author} {\bibfnamefont {A.}~\bibnamefont {Meinsma}}, \bibinfo {author} {\bibfnamefont {M.}~\bibnamefont {Markham}}, \bibinfo {author} {\bibfnamefont {D.}~\bibnamefont {Twitchen}}, \ and\ \bibinfo {author} {\bibfnamefont {T.}~\bibnamefont {Taminiau}},\ }\href@noop {} {\bibfield  {journal} {\bibinfo  {journal} {Nature Communications}\ }\textbf {\bibinfo {volume} {12}},\ \bibinfo {pages} {3470} (\bibinfo {year} {2021})}\BibitemShut {NoStop}%
\bibitem [{\citenamefont {Ungar}\ \emph {et~al.}(2024)\citenamefont {Ungar}, \citenamefont {Cappellaro}, \citenamefont {Cooper},\ and\ \citenamefont {Sun}}]{ungar2024control}%
  \BibitemOpen
  \bibfield  {author} {\bibinfo {author} {\bibfnamefont {A.}~\bibnamefont {Ungar}}, \bibinfo {author} {\bibfnamefont {P.}~\bibnamefont {Cappellaro}}, \bibinfo {author} {\bibfnamefont {A.}~\bibnamefont {Cooper}}, \ and\ \bibinfo {author} {\bibfnamefont {W.~K.~C.}\ \bibnamefont {Sun}},\ }\href@noop {} {\bibfield  {journal} {\bibinfo  {journal} {PRX Quantum}\ }\textbf {\bibinfo {volume} {5}},\ \bibinfo {pages} {010321} (\bibinfo {year} {2024})}\BibitemShut {NoStop}%
\bibitem [{\citenamefont {Topuz}\ \emph {et~al.}(2013)\citenamefont {Topuz}, \citenamefont {G{\"u}nd{\"u}z}, \citenamefont {Mavis},\ and\ \citenamefont {{\c{C}}olak}}]{topuz2013synthesis}%
  \BibitemOpen
  \bibfield  {author} {\bibinfo {author} {\bibfnamefont {B.~B.}\ \bibnamefont {Topuz}}, \bibinfo {author} {\bibfnamefont {G.}~\bibnamefont {G{\"u}nd{\"u}z}}, \bibinfo {author} {\bibfnamefont {B.}~\bibnamefont {Mavis}}, \ and\ \bibinfo {author} {\bibfnamefont {{\"U}.}~\bibnamefont {{\c{C}}olak}},\ }\href@noop {} {\bibfield  {journal} {\bibinfo  {journal} {Dyes and Pigments}\ }\textbf {\bibinfo {volume} {96}},\ \bibinfo {pages} {31} (\bibinfo {year} {2013})}\BibitemShut {NoStop}%
\bibitem [{\citenamefont {Cranston}\ and\ \citenamefont {Lessard}(2021)}]{cranston2021metal}%
  \BibitemOpen
  \bibfield  {author} {\bibinfo {author} {\bibfnamefont {R.~R.}\ \bibnamefont {Cranston}}\ and\ \bibinfo {author} {\bibfnamefont {B.~H.}\ \bibnamefont {Lessard}},\ }\href@noop {} {\bibfield  {journal} {\bibinfo  {journal} {RSC advances}\ }\textbf {\bibinfo {volume} {11}},\ \bibinfo {pages} {21716} (\bibinfo {year} {2021})}\BibitemShut {NoStop}%
\bibitem [{\citenamefont {Steinert}\ \emph {et~al.}(2013)\citenamefont {Steinert}, \citenamefont {Ziem}, \citenamefont {Hall}, \citenamefont {Zappe}, \citenamefont {Schweikert}, \citenamefont {G{\"o}tz}, \citenamefont {Aird}, \citenamefont {Balasubramanian}, \citenamefont {Hollenberg},\ and\ \citenamefont {Wrachtrup}}]{steinert2013magnetic}%
  \BibitemOpen
  \bibfield  {author} {\bibinfo {author} {\bibfnamefont {S.}~\bibnamefont {Steinert}}, \bibinfo {author} {\bibfnamefont {F.}~\bibnamefont {Ziem}}, \bibinfo {author} {\bibfnamefont {L.}~\bibnamefont {Hall}}, \bibinfo {author} {\bibfnamefont {A.}~\bibnamefont {Zappe}}, \bibinfo {author} {\bibfnamefont {M.}~\bibnamefont {Schweikert}}, \bibinfo {author} {\bibfnamefont {N.}~\bibnamefont {G{\"o}tz}}, \bibinfo {author} {\bibfnamefont {A.}~\bibnamefont {Aird}}, \bibinfo {author} {\bibfnamefont {G.}~\bibnamefont {Balasubramanian}}, \bibinfo {author} {\bibfnamefont {L.}~\bibnamefont {Hollenberg}}, \ and\ \bibinfo {author} {\bibfnamefont {J.}~\bibnamefont {Wrachtrup}},\ }\href@noop {} {\bibfield  {journal} {\bibinfo  {journal} {Nature communications}\ }\textbf {\bibinfo {volume} {4}},\ \bibinfo {pages} {1607} (\bibinfo {year} {2013})}\BibitemShut {NoStop}%
\bibitem [{\citenamefont {Pelliccione}\ \emph {et~al.}(2014)\citenamefont {Pelliccione}, \citenamefont {Myers}, \citenamefont {Pascal}, \citenamefont {Das},\ and\ \citenamefont {Bleszynski~Jayich}}]{pelliccione2014two}%
  \BibitemOpen
  \bibfield  {author} {\bibinfo {author} {\bibfnamefont {M.}~\bibnamefont {Pelliccione}}, \bibinfo {author} {\bibfnamefont {B.~A.}\ \bibnamefont {Myers}}, \bibinfo {author} {\bibfnamefont {L.}~\bibnamefont {Pascal}}, \bibinfo {author} {\bibfnamefont {A.}~\bibnamefont {Das}}, \ and\ \bibinfo {author} {\bibfnamefont {A.}~\bibnamefont {Bleszynski~Jayich}},\ }\href@noop {} {\bibfield  {journal} {\bibinfo  {journal} {Physical Review Applied}\ }\textbf {\bibinfo {volume} {2}},\ \bibinfo {pages} {054014} (\bibinfo {year} {2014})}\BibitemShut {NoStop}%
\bibitem [{\citenamefont {Kumar}\ \emph {et~al.}(2024)\citenamefont {Kumar}, \citenamefont {Yudilevich}, \citenamefont {Smooha}, \citenamefont {Zohar}, \citenamefont {Pariari}, \citenamefont {St\"{o}hr}, \citenamefont {Denisenko}, \citenamefont {H\"{u}cker},\ and\ \citenamefont {Finkler}}]{kumar2024room}%
  \BibitemOpen
  \bibfield  {author} {\bibinfo {author} {\bibfnamefont {J.}~\bibnamefont {Kumar}}, \bibinfo {author} {\bibfnamefont {D.}~\bibnamefont {Yudilevich}}, \bibinfo {author} {\bibfnamefont {A.}~\bibnamefont {Smooha}}, \bibinfo {author} {\bibfnamefont {I.}~\bibnamefont {Zohar}}, \bibinfo {author} {\bibfnamefont {A.~K.}\ \bibnamefont {Pariari}}, \bibinfo {author} {\bibfnamefont {R.}~\bibnamefont {St\"{o}hr}}, \bibinfo {author} {\bibfnamefont {A.}~\bibnamefont {Denisenko}}, \bibinfo {author} {\bibfnamefont {M.}~\bibnamefont {H\"{u}cker}}, \ and\ \bibinfo {author} {\bibfnamefont {A.}~\bibnamefont {Finkler}},\ }\href@noop {} {\bibfield  {journal} {\bibinfo  {journal} {Nano Letters}\ }\textbf {\bibinfo {volume} {24}},\ \bibinfo {pages} {4793} (\bibinfo {year} {2024})}\BibitemShut {NoStop}%
\bibitem [{\citenamefont {Li}\ \emph {et~al.}(2019)\citenamefont {Li}, \citenamefont {Chen}, \citenamefont {Lyzwa},\ and\ \citenamefont {Cappellaro}}]{li2019all}%
  \BibitemOpen
  \bibfield  {author} {\bibinfo {author} {\bibfnamefont {C.}~\bibnamefont {Li}}, \bibinfo {author} {\bibfnamefont {M.}~\bibnamefont {Chen}}, \bibinfo {author} {\bibfnamefont {D.}~\bibnamefont {Lyzwa}}, \ and\ \bibinfo {author} {\bibfnamefont {P.}~\bibnamefont {Cappellaro}},\ }\href@noop {} {\bibfield  {journal} {\bibinfo  {journal} {Nano Letters}\ }\textbf {\bibinfo {volume} {19}},\ \bibinfo {pages} {7342} (\bibinfo {year} {2019})}\BibitemShut {NoStop}%
\bibitem [{\citenamefont {Grinolds}\ \emph {et~al.}(2014)\citenamefont {Grinolds}, \citenamefont {Warner}, \citenamefont {De~Greve}, \citenamefont {Dovzhenko}, \citenamefont {Thiel}, \citenamefont {Walsworth}, \citenamefont {Hong}, \citenamefont {Maletinsky},\ and\ \citenamefont {Yacoby}}]{grinolds2014subnanometre}%
  \BibitemOpen
  \bibfield  {author} {\bibinfo {author} {\bibfnamefont {M.}~\bibnamefont {Grinolds}}, \bibinfo {author} {\bibfnamefont {M.}~\bibnamefont {Warner}}, \bibinfo {author} {\bibfnamefont {K.}~\bibnamefont {De~Greve}}, \bibinfo {author} {\bibfnamefont {Y.}~\bibnamefont {Dovzhenko}}, \bibinfo {author} {\bibfnamefont {L.}~\bibnamefont {Thiel}}, \bibinfo {author} {\bibfnamefont {R.~L.}\ \bibnamefont {Walsworth}}, \bibinfo {author} {\bibfnamefont {S.}~\bibnamefont {Hong}}, \bibinfo {author} {\bibfnamefont {P.}~\bibnamefont {Maletinsky}}, \ and\ \bibinfo {author} {\bibfnamefont {A.}~\bibnamefont {Yacoby}},\ }\href@noop {} {\bibfield  {journal} {\bibinfo  {journal} {Nature nanotechnology}\ }\textbf {\bibinfo {volume} {9}},\ \bibinfo {pages} {279} (\bibinfo {year} {2014})}\BibitemShut {NoStop}%
\bibitem [{\citenamefont {Shaibat}\ \emph {et~al.}(2010)\citenamefont {Shaibat}, \citenamefont {Casabianca}, \citenamefont {Siberio-P{\'e}rez}, \citenamefont {Matzger},\ and\ \citenamefont {Ishii}}]{shaibat2010distinguishing}%
  \BibitemOpen
  \bibfield  {author} {\bibinfo {author} {\bibfnamefont {M.~A.}\ \bibnamefont {Shaibat}}, \bibinfo {author} {\bibfnamefont {L.~B.}\ \bibnamefont {Casabianca}}, \bibinfo {author} {\bibfnamefont {D.~Y.}\ \bibnamefont {Siberio-P{\'e}rez}}, \bibinfo {author} {\bibfnamefont {A.~J.}\ \bibnamefont {Matzger}}, \ and\ \bibinfo {author} {\bibfnamefont {Y.}~\bibnamefont {Ishii}},\ }\href@noop {} {\bibfield  {journal} {\bibinfo  {journal} {The Journal of Physical Chemistry B}\ }\textbf {\bibinfo {volume} {114}},\ \bibinfo {pages} {4400} (\bibinfo {year} {2010})}\BibitemShut {NoStop}%
\bibitem [{\citenamefont {Choi}\ \emph {et~al.}(2017)\citenamefont {Choi}, \citenamefont {Choi}, \citenamefont {Kucsko}, \citenamefont {Maurer}, \citenamefont {Shields}, \citenamefont {Sumiya}, \citenamefont {Onoda}, \citenamefont {Isoya}, \citenamefont {Demler}, \citenamefont {Jelezko} \emph {et~al.}}]{choi2017depolarization}%
  \BibitemOpen
  \bibfield  {author} {\bibinfo {author} {\bibfnamefont {J.}~\bibnamefont {Choi}}, \bibinfo {author} {\bibfnamefont {S.}~\bibnamefont {Choi}}, \bibinfo {author} {\bibfnamefont {G.}~\bibnamefont {Kucsko}}, \bibinfo {author} {\bibfnamefont {P.~C.}\ \bibnamefont {Maurer}}, \bibinfo {author} {\bibfnamefont {B.~J.}\ \bibnamefont {Shields}}, \bibinfo {author} {\bibfnamefont {H.}~\bibnamefont {Sumiya}}, \bibinfo {author} {\bibfnamefont {S.}~\bibnamefont {Onoda}}, \bibinfo {author} {\bibfnamefont {J.}~\bibnamefont {Isoya}}, \bibinfo {author} {\bibfnamefont {E.}~\bibnamefont {Demler}}, \bibinfo {author} {\bibfnamefont {F.}~\bibnamefont {Jelezko}},  \emph {et~al.},\ }\href@noop {} {\bibfield  {journal} {\bibinfo  {journal} {Physical review letters}\ }\textbf {\bibinfo {volume} {118}},\ \bibinfo {pages} {093601} (\bibinfo {year} {2017})}\BibitemShut {NoStop}%
\bibitem [{\citenamefont {Machado}\ \emph {et~al.}(2023)\citenamefont {Machado}, \citenamefont {Demler}, \citenamefont {Yao},\ and\ \citenamefont {Chatterjee}}]{machado2023quantum}%
  \BibitemOpen
  \bibfield  {author} {\bibinfo {author} {\bibfnamefont {F.}~\bibnamefont {Machado}}, \bibinfo {author} {\bibfnamefont {E.~A.}\ \bibnamefont {Demler}}, \bibinfo {author} {\bibfnamefont {N.~Y.}\ \bibnamefont {Yao}}, \ and\ \bibinfo {author} {\bibfnamefont {S.}~\bibnamefont {Chatterjee}},\ }\href@noop {} {\bibfield  {journal} {\bibinfo  {journal} {Physical Review Letters}\ }\textbf {\bibinfo {volume} {131}},\ \bibinfo {pages} {070801} (\bibinfo {year} {2023})}\BibitemShut {NoStop}%
\bibitem [{\citenamefont {Slichter}(2013)}]{slichter2013principles}%
  \BibitemOpen
  \bibfield  {author} {\bibinfo {author} {\bibfnamefont {C.~P.}\ \bibnamefont {Slichter}},\ }\href@noop {} {\emph {\bibinfo {title} {Principles of magnetic resonance}}},\ Vol.~\bibinfo {volume} {1}\ (\bibinfo  {publisher} {Springer Science \& Business Media},\ \bibinfo {year} {2013})\BibitemShut {NoStop}%
\bibitem [{smo()}]{smofthispaper}%
  \BibitemOpen
  \href@noop {} {\bibinfo  {journal} {Supplemental Materials: Quantum Sensing of Copper-Phthalocyanine Electron Spins via NV Relaxometry}\ }\BibitemShut {NoStop}%
\bibitem [{\citenamefont {Rodriguez-Nieva}\ \emph {et~al.}(2018)\citenamefont {Rodriguez-Nieva}, \citenamefont {Agarwal}, \citenamefont {Giamarchi}, \citenamefont {Halperin}, \citenamefont {Lukin},\ and\ \citenamefont {Demler}}]{rodriguez2018probing}%
  \BibitemOpen
\bibfield  {journal} {  }\bibfield  {author} {\bibinfo {author} {\bibfnamefont {J.~F.}\ \bibnamefont {Rodriguez-Nieva}}, \bibinfo {author} {\bibfnamefont {K.}~\bibnamefont {Agarwal}}, \bibinfo {author} {\bibfnamefont {T.}~\bibnamefont {Giamarchi}}, \bibinfo {author} {\bibfnamefont {B.~I.}\ \bibnamefont {Halperin}}, \bibinfo {author} {\bibfnamefont {M.~D.}\ \bibnamefont {Lukin}}, \ and\ \bibinfo {author} {\bibfnamefont {E.}~\bibnamefont {Demler}},\ }\href@noop {} {\bibfield  {journal} {\bibinfo  {journal} {Physical Review B}\ }\textbf {\bibinfo {volume} {98}},\ \bibinfo {pages} {195433} (\bibinfo {year} {2018})}\BibitemShut {NoStop}%
\bibitem [{\citenamefont {Finazzo}\ \emph {et~al.}(2006)\citenamefont {Finazzo}, \citenamefont {Calle}, \citenamefont {Stoll}, \citenamefont {Van~Doorslaer},\ and\ \citenamefont {Schweiger}}]{finazzo2006matrix}%
  \BibitemOpen
  \bibfield  {author} {\bibinfo {author} {\bibfnamefont {C.}~\bibnamefont {Finazzo}}, \bibinfo {author} {\bibfnamefont {C.}~\bibnamefont {Calle}}, \bibinfo {author} {\bibfnamefont {S.}~\bibnamefont {Stoll}}, \bibinfo {author} {\bibfnamefont {S.}~\bibnamefont {Van~Doorslaer}}, \ and\ \bibinfo {author} {\bibfnamefont {A.}~\bibnamefont {Schweiger}},\ }\href@noop {} {\bibfield  {journal} {\bibinfo  {journal} {Physical Chemistry Chemical Physics}\ }\textbf {\bibinfo {volume} {8}},\ \bibinfo {pages} {1942} (\bibinfo {year} {2006})}\BibitemShut {NoStop}%
\bibitem [{\citenamefont {Barbon}\ \emph {et~al.}(2001)\citenamefont {Barbon}, \citenamefont {Brustolon},\ and\ \citenamefont {van Faassen}}]{barbon2001photoexcited}%
  \BibitemOpen
  \bibfield  {author} {\bibinfo {author} {\bibfnamefont {A.}~\bibnamefont {Barbon}}, \bibinfo {author} {\bibfnamefont {M.}~\bibnamefont {Brustolon}}, \ and\ \bibinfo {author} {\bibfnamefont {E.~E.}\ \bibnamefont {van Faassen}},\ }\href@noop {} {\bibfield  {journal} {\bibinfo  {journal} {Physical Chemistry Chemical Physics}\ }\textbf {\bibinfo {volume} {3}},\ \bibinfo {pages} {5342} (\bibinfo {year} {2001})}\BibitemShut {NoStop}%
\bibitem [{\citenamefont {Ho}\ and\ \citenamefont {Chibotaru}(2018)}]{ho2018spin}%
  \BibitemOpen
  \bibfield  {author} {\bibinfo {author} {\bibfnamefont {L.~T.~A.}\ \bibnamefont {Ho}}\ and\ \bibinfo {author} {\bibfnamefont {L.~F.}\ \bibnamefont {Chibotaru}},\ }\href@noop {} {\bibfield  {journal} {\bibinfo  {journal} {Physical Review B}\ }\textbf {\bibinfo {volume} {97}},\ \bibinfo {pages} {024427} (\bibinfo {year} {2018})}\BibitemShut {NoStop}%
\bibitem [{\citenamefont {Zhou}\ \emph {et~al.}(2024)\citenamefont {Zhou}, \citenamefont {Li}, \citenamefont {Tan}, \citenamefont {Lv}, \citenamefont {Pang}, \citenamefont {Zhao}, \citenamefont {Shi}, \citenamefont {Zhang}, \citenamefont {Jin}, \citenamefont {Liu} \emph {et~al.}}]{zhou2024phononic}%
  \BibitemOpen
  \bibfield  {author} {\bibinfo {author} {\bibfnamefont {A.}~\bibnamefont {Zhou}}, \bibinfo {author} {\bibfnamefont {D.}~\bibnamefont {Li}}, \bibinfo {author} {\bibfnamefont {M.}~\bibnamefont {Tan}}, \bibinfo {author} {\bibfnamefont {Y.}~\bibnamefont {Lv}}, \bibinfo {author} {\bibfnamefont {S.}~\bibnamefont {Pang}}, \bibinfo {author} {\bibfnamefont {X.}~\bibnamefont {Zhao}}, \bibinfo {author} {\bibfnamefont {Z.}~\bibnamefont {Shi}}, \bibinfo {author} {\bibfnamefont {J.}~\bibnamefont {Zhang}}, \bibinfo {author} {\bibfnamefont {F.}~\bibnamefont {Jin}}, \bibinfo {author} {\bibfnamefont {S.}~\bibnamefont {Liu}},  \emph {et~al.},\ }\href@noop {} {\bibfield  {journal} {\bibinfo  {journal} {Nature Communications}\ }\textbf {\bibinfo {volume} {15}},\ \bibinfo {pages} {10763} (\bibinfo {year} {2024})}\BibitemShut {NoStop}%
\bibitem [{\citenamefont {Li}\ and\ \citenamefont {et~al.}()}]{li25eprpaper}%
  \BibitemOpen
  \bibfield  {author} {\bibinfo {author} {\bibfnamefont {B.}~\bibnamefont {Li}}\ and\ \bibinfo {author} {\bibnamefont {et~al.}},\ }\href@noop {} {\enquote {\bibinfo {title} {Exploring the mechanisms of transverse relaxation of copper(ii)-phthalocyanine spin qubits},}\ }\bibinfo {note} {Companion paper, submitted to arXiv (2025)}\BibitemShut {NoStop}%
\bibitem [{\citenamefont {Stone}(2019)}]{stone2019table}%
  \BibitemOpen
  \bibfield  {author} {\bibinfo {author} {\bibfnamefont {N.}~\bibnamefont {Stone}},\ }\href@noop {} {\emph {\bibinfo {title} {Table of recommended nuclear magnetic dipole moments}}},\ \bibinfo {type} {Tech. Rep.}\ (\bibinfo  {institution} {International Atomic Energy Agency},\ \bibinfo {year} {2019})\BibitemShut {NoStop}%
\bibitem [{\citenamefont {Cohen-Tannoudji}\ \emph {et~al.}(1998)\citenamefont {Cohen-Tannoudji}, \citenamefont {Dupont-Roc},\ and\ \citenamefont {Grynberg}}]{cohen1998atom}%
  \BibitemOpen
  \bibfield  {author} {\bibinfo {author} {\bibfnamefont {C.}~\bibnamefont {Cohen-Tannoudji}}, \bibinfo {author} {\bibfnamefont {J.}~\bibnamefont {Dupont-Roc}}, \ and\ \bibinfo {author} {\bibfnamefont {G.}~\bibnamefont {Grynberg}},\ }\href@noop {} {\emph {\bibinfo {title} {Atom-photon interactions: basic processes and applications}}}\ (\bibinfo  {publisher} {John Wiley \& Sons},\ \bibinfo {year} {1998})\BibitemShut {NoStop}%
\bibitem [{\citenamefont {Luis}\ and\ \citenamefont {Fern{\'a}ndez}(2014)}]{luis2014molecular}%
  \BibitemOpen
  \bibfield  {author} {\bibinfo {author} {\bibfnamefont {F.}~\bibnamefont {Luis}}\ and\ \bibinfo {author} {\bibfnamefont {J.~F.}\ \bibnamefont {Fern{\'a}ndez}},\ }\href@noop {} {\emph {\bibinfo {title} {Molecular Magnets: Physics and Applications}}}\ (\bibinfo  {publisher} {Imprint: Springer},\ \bibinfo {year} {2014})\BibitemShut {NoStop}%
\end{thebibliography}%

\end{document}

\begin{center}
{\bf \large{Supplemental Materials:}}\\
{\bf \Large{Quantum Sensing of Copper-Phthalocyanine Electron Spins via NV Relaxometry}}
\end{center}
% \iffalse

\author{Boning Li}\thanks{These authors contributed equally.}
    \affiliation{Department of Physics, Massachusetts Institute of Technology, Cambridge, MA 02139, USA}
    \affiliation{Research Laboratory of Electronics, Massachusetts Institute of Technology, Cambridge, MA 02139, USA}
\author{Xufan Li}\thanks{These authors contributed equally.}
   \affiliation{Honda Research Institute USA, Inc., San Jose, CA 95134, USA}
\author{Yifan Quan}
    \affiliation{Department of Chemistry and Francis Bitter Magnet Laboratory, Massachusetts Institute of Technology, Cambridge, MA 02139, USA}
\author{Avetik R Harutyunyan}\email{aharutyunyan@honda-ri.com}
    \affiliation{Honda Research Institute USA, Inc., San Jose, CA 95134, USA}
    \affiliation{Department of Nuclear Science and Engineering, Massachusetts Institute of Technology, Cambridge, MA 02139, USA}
\author{Paola Cappellaro}\email[]{pcappell@mit.edu}
    \affiliation{Department of Physics, Massachusetts Institute of Technology, Cambridge, MA 02139, USA}
    \affiliation{Research Laboratory of Electronics, Massachusetts Institute of Technology, Cambridge, MA 02139, USA}
    \affiliation{Department of Nuclear Science and Engineering, Massachusetts Institute of Technology, Cambridge, MA 02139, USA}
% \fi

\maketitle

\tableofcontents

\section{Experiment setup}
Our experiment was conducted using a home-built confocal microscope. 532nm laser (SPROUT from Lighthouse Photonics) beam is sent through an acousto-optic modulator (AOM, Isomet Corporation, M113-aQ80L-H) with fast switching and then focused using an oil immersion objective (Thorlabs N100X-PFO Nikon Plan Flour). The photon illuminance is collected using the same objective and separated from the excitation light using a diachoric mirror (Chroma NC338988). Filtered by a 549nm long pass and a 715nm short pass filter, the fluorescence is collected using a single-photon counting module (Perkin ElmerSPCM-AQRH-14). The magnetic field is applied from a composite assembly of 12 N52 cubic magnets of 9.53 mm edge (K$\&$J Magnetics B666-N52).

The T1 experiment conducted in the maintext is microwave free, including the magnetic field alignment which is based on NV photoluminescence quenching\cite{liu2019nanoscale}. However, to employing ODMR and other dynamical decoupling experiments, microwave is delivered to the sample using a coplanar wave guide. Microwave pulse is programmed using an arbitrary wave generator (Techtronics AWG5014C) with pre-amplifier (ZHL-30W-252-S+).
\section{Diamond sample preparation and characterization}
\begin{figure}[htbp]
\centering
\includegraphics[width=\linewidth]{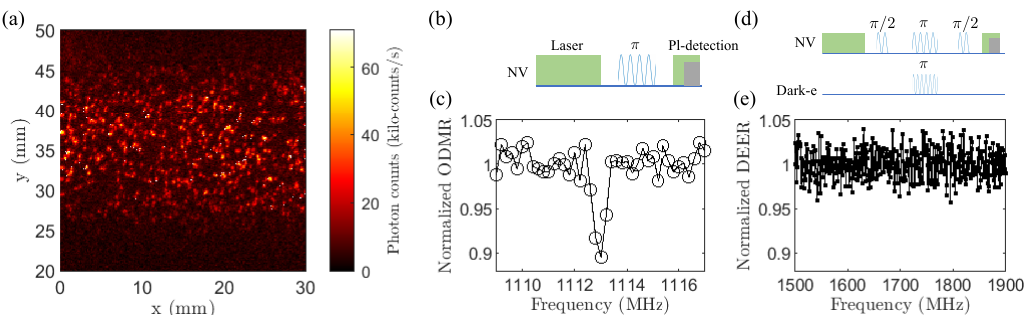}
\caption{
NV center characterization.  
(a) Fluorescence image of NV centers in the diamond prior to CuPc deposition. Bright spots correspond to individual NV centers.  
(b) Pulse sequence used for optically detected magnetic resonance (ODMR).  
(c) ODMR spectrum of a single NV center under an external magnetic field of 621~G aligned along the NV axis, showing a single resonance peak corresponding to the $|0\rangle \rightarrow |-1\rangle$ transition.  
(d) Pulse sequence for double electron–electron resonance (DEER).  
(e) DEER measurement results, indicating the absence of electron spin environment around the NV center.
}

\label{figS:diamond}
\end{figure}

Our experimental sample consists of an isotopically pure diamond (99.99$\%$ $^{12}$C, Element six) containing NV centers produced via
 implantation ($^{15}$N, Innovion) and subsequent annealing \cite{sangtawesin2019origins}. An optical image of shallow NV centers on bare diamond is hown in Fig.S\ref{figS:diamond} by measuring the florescence of NV center of larger than 715nm wavelength.

 We conducted double-electron-electron-resonance (DEER) experiment to characterize the surface electron spin reported on previous shallow NV center research \cite{sushkov2014magnetic}. We tested on various NV centers at magnetic field $\sim$490G, no signal of additional electron spins was observed. This is result from the annealing process and the termination change of the diamond that removes additional spin and charge on the surface. This confirms that our diamond has a uniform electron-free surface and the additional depolarization we observed in the main text in from the additional electron spin layer, CuPc.
\section{CuPc thin layer deposition}
\subsection{Characterization}
\begin{figure}[htbp]
\centering
\includegraphics[width=\linewidth]{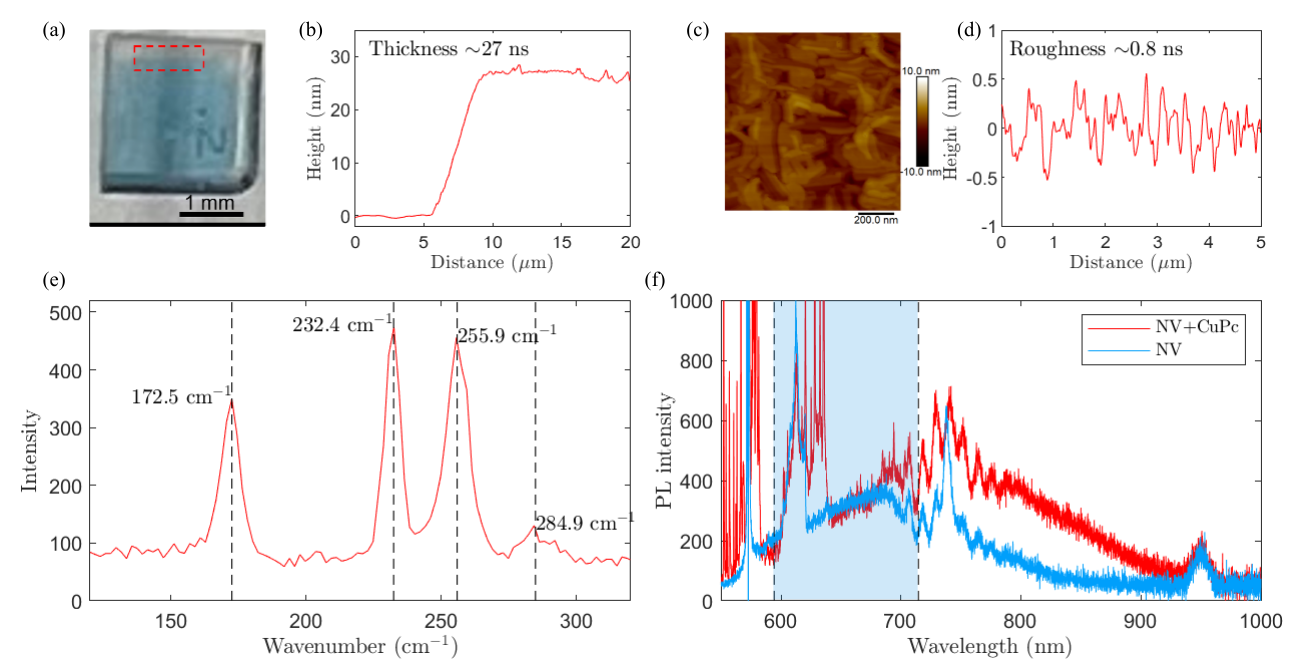}
\caption{ CuPc layer characterization.
(a) Optical image of a diamond coated with a CuPc thin film (blue area). (b) The thickness of the CuPc layer, measured by atomic force microscopy (AFM) from the regions containing the interface between the film and the bare diamond surface (i.e., indicated by the dashed red box in (a)). The height differences across the interfaces show that the thickness of the CuPc film is $\sim$27 nm. (c) AFM image showing the microscopic morphology of the CuPc film. (d) Surface roughness of the CuPc film measured by AFM, which is $\sim$0.8 nm. 
(e) Raman spectra of the CuPc film deposited on NV-diamond, measured with 532 nm laser excitation. Characteristic peaks at 172.5 cm-1, 232.4 cm-1, 255.9 cm-1, and 284.9 cm-1 confirms that the deposited CuPc crystal is alpha-phase. (f) Photoluminescence (PL) spectrum (red curve) of the CuPc film deposited on NV-diamond under 532 nm laser excitation. A PL spectra of NV center is also presented (blue curve) for comparison. In the 594 nm to 715 nm range (shaded area), PL is dominated by NV centers.  
}
\label{figS:cupc}
\end{figure}

This is the first time of depositing thin film of copper phthalocyanine onto the surface of diamond. The deposition was conducted by thermal evaporation of CuPc powder (Millipore Sigma, $>99\%$) in a physical vapor deposition system (nanoPVD-T15A, Moorfield Nanotechnology) under a vacuum of $\sim1\times10^{-6} \text{Torr}$. The as purchased CuPc powder was further purified through sublimation before deposition. The CuPc was deposited on diamond at a rate of $\sim$0.1 nm/s at which the CuPc powder was heated to $\sim$310 \degree C, while the diamond mounted on the substrate holder was heated to $\sim$250 \degree C. The thickness of the as-deposited CuPc film is $\sim$27 nm, with a surface roughness of $\sim$0.8 nm, as confirmed by AFM analysis, shown in Fig.S\ref{figS:cupc}.

We conducted Raman experiment to determine the the thin-film crystal type and also the fluorescence effects on NV center. As shown in Fig.S\ref{figS:cupc} (b). With 532nm excitation, characteristic peaks at 172.5 cm-1, 232.4 cm-1, 255.9 cm-1, and 284.9 cm-1 confirms that the deposited CuPc crystal is alpha-phase \cite{shaibat2010distinguishing}. The fluorescence spectra of NV and CuPc composite shown in Fig.S\ref{figS:cupc} (f) suggests the two spectra are distinct enough in frequency space so that they could both be detected. In particular, we can still see, and filter out using an appropriate filter (715 nm short pass filter). Still, comparing a confocal image of a pristine diamond surface (as found on the edge of the diamond) and a portion of the diamond covered by the CuPc film, we notice that the CuPc contributes to added background noise. Still, the brightest NV emission can be seen with about the same count rate.

\subsection{Stability of the CuPc thin film under laser illumination}
We further evaluated whether the film could withstand the laser illumination needed for a typical experiment. We considered two types of experiments: imaging and optically detected magnetic resonance (ODMR). The first is used to scan the sample and identify NV centers with laser power 0.5~mW. Here, for each pixel the dwell time is 0.0096s with 200 averages (total integration time is 0.0096$\times$200 $=$ 1.92~s) over the field of view. For ODMR, the exact sequence time depends on the length of the NV spin evolution in the dark, which varies from a few microseconds for Ramsey to potentially milliseconds for T1 relaxation. For each data point, we use a combined 5.5~${\mu}$s for optical initialization and readout. We also take a reference readout to normalize the signal in the presence of technical fluctuations. In order to obtain a good signal-to-noise ratio (since CuPc induces a larger photon background noise) we use 5$\times$10$^5$ repetitions for each point. Each curve is averaged 10 times. We often perform tracking in between averages, that is, perform a local confocal scan to find the optimal sample position at the NV spot, to account for micromotions of the setup. Neglecting this tracking, the laser illumination time is about 165~s (it becomes about 1800~s with tracking). We note that this time is not continuous, thus heating effects are reduced, especially in light of the high thermal conductivity of diamond. To verify that this amount of laser illumination would not damage the CuPc film, we illuminated a single spot with a higher laser power (1.1~mW instead of 0.5~mW as in experiments). Only after illumination of about 10 minutes we see that the laser damages (burns out) the CuPc. Typical experimental illumination does not.

This burnt of CuPc is also observed using optical microscope to see dark areas under regular LED illuminance. The damage if the thin film confirms it is indeed CuPc removal instead of transforming into dark materials, shown in Fig. S\ref{figS:cupc_burn}. As CuPc contributes additional fluorescence as a background that reduce the contrast of ODMR of NV center, we can use the ODMR contrast of NV center to monitor the degree of burnt of CuPc on top of the NV center after keep illuminating with high power laser for some time. We fit the ODMR data with Lorentzian function and extract the peak height as contrast, which increases as laser time increases and saturates until the CuPc is fully burnt locally. 
Here we observe that the maximum contrast ($\sim$10\%) recovers to the same level as the CuPc-free case after 600~s of illumination, indicating successful removal of the CuPc layer.
%still smaller than the typical contrast of bare shallow NV centers ($\sim$15$\%$) on the same sample which is likely due to random reflection and fluorescence of non-burnt CuPc from surface.
 \begin{figure}[htbp]
\centering
\includegraphics[width=\linewidth]{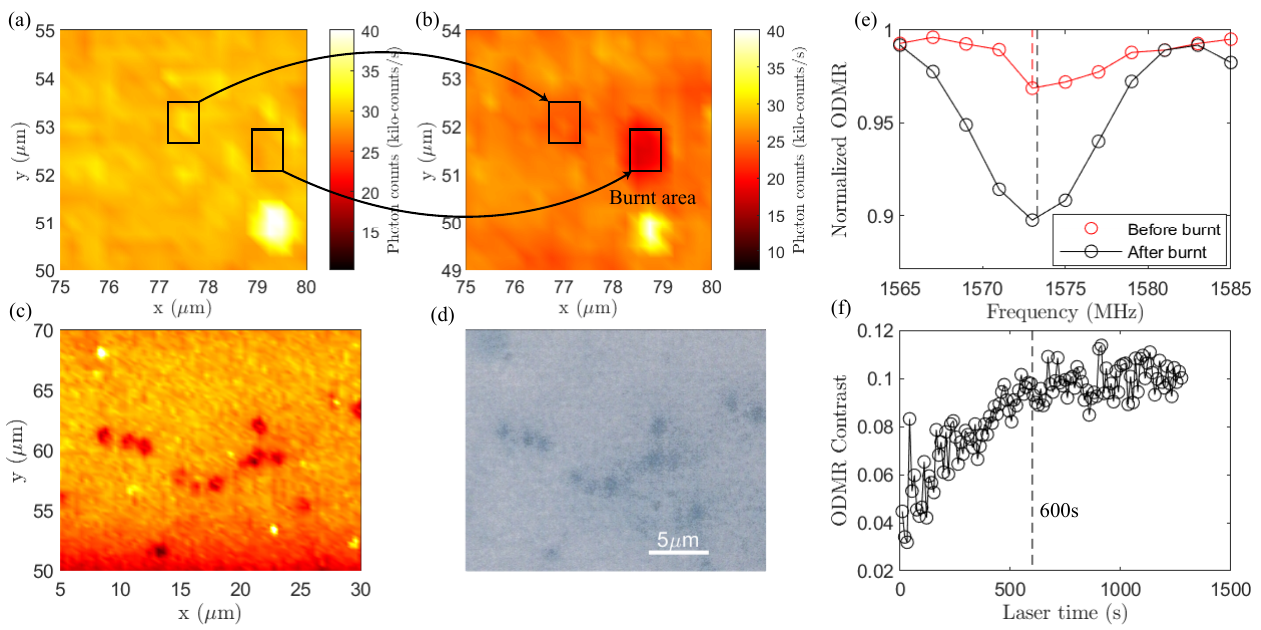}
\caption{Confocal images of NV centers before (left) and after (right) longer optical illumination. Only the
 spot illuminated for 10 minutes shows damage, while other spots used for ODMR are untouched. 
}
\label{figS:cupc_burn}
\end{figure}

\begin{figure}[htbp]
\centering
\includegraphics[width=\linewidth]{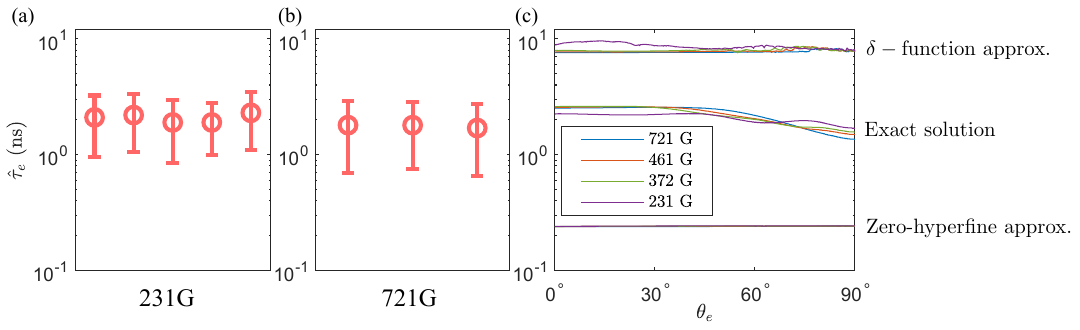}
\caption{
CuPc correlation time estimation.  
(a,b) Experimentally extracted correlation times $\tau_e$ of CuPc for individual NV centers measured at 231~G and 721~G, respectively. Consistent values across different NV centers indicate good uniformity of the CuPc layer.  
(c) Numerically calculated correlation times of CuPc as a function of the angle between the molecular axis and the external magnetic field (as defined in the main text). The three groups correspond to different approximations used for the spin bath spectral density. Each group includes four curves representing the four magnetic field strengths used in the experiments. Experiment result shows consistency with the result of the exact solution of the self-consistent function derived in Appendix.~B.
}
\label{figS:cupc_correlation_calculation}
\end{figure}
\begin{figure}[htbp]
\centering
\includegraphics[width=\linewidth]{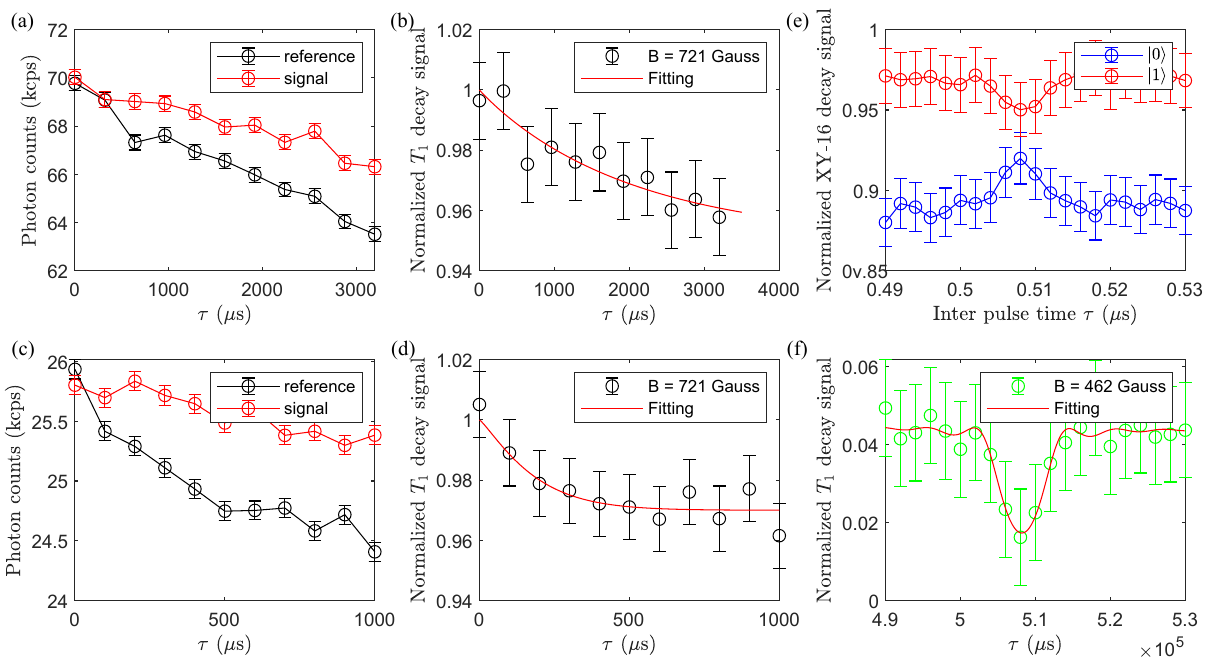}
\caption{$T_1$ and depth measurement results. 
(a, b) Depolarization signal of the NV center measured after CuPc removal using high-power laser illumination. 
(b) shows the normalized signal (signal/reference from (a)) fitted with the function $y = A \exp\left[-(t/T_1)^{\iota}\right] + C$. 
(c, d) Depolarization signal of the NV center measured with the CuPc layer present. 
(e, f) NV center depth estimation using the method described in Ref.~\cite{pham2016nmr}. 
(f) presents the normalized data with fitted curve.
}

\label{figS:cupc_burn}
\end{figure}

To infer the T1 time at room temperature due to spin-lattice interaction, we fit current data with expression that captures the contribution from one- and two- photon to the spin life time in vdW crystals of magnetic molecules \cite{lunghi2022toward}, as well as a constant term for the invariant contribute from nuclear spins:
\begin{equation}
    \frac{1}{T_{s-p}} = \frac{A}{e^{\frac{\omega_A}{k_BT}}-1}+\frac{Be^{\frac{\omega_B}{k_BT}}}{(e^{\frac{\omega_B}{k_BT}}-1)^2} +C,
\end{equation}
where $k_B$ is the Boltzmann constant, and we added  At our experiment condition at room temperature ($T = 300K$), the depolarization rate is estimated to be $\frac{1}{T_{s-p}} \approx \frac{1}{38}\ \text{ns}^{-1}$, as we used in the maintext. A detailed study of CuPc coherence properties is shown in another work~\cite{li25eprpaper}.
 \begin{figure}[htbp]
\centering
\includegraphics[width=0.7\linewidth]{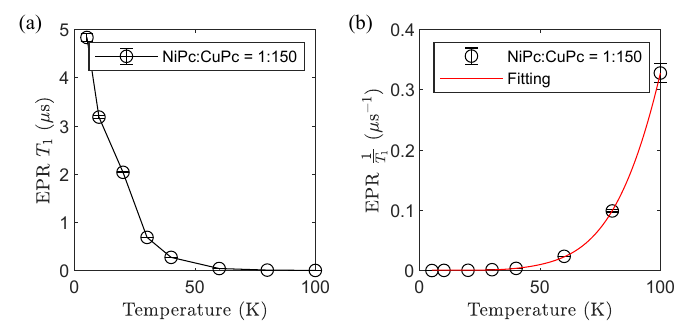}
\caption{CuPc T1 epr.
}
\label{figS:cupc_epr}
\end{figure}
\bibliography{CuPc_NV_ref}% Produces the bibliography via BibTeX.